\documentclass[twocolumn]{aa}
\usepackage{graphicx}
\begin{document}

\def\ie{{\it i.e.}}
\def\kms{~km~s$^{-1}$}
\def\micron{$\mu$m}
\def\fe2{[Fe\,{\sc ii}]}
\def\h2{H$_{2}$}

   \title{Molecular Hydrogen and [\ion{Fe}{ii}] in Active Galactic Nuclei}

   \subtitle{}

\titlerunning{H$_{2}$ and [\ion{Fe}{ii}] in AGN}

   \author{Alberto Rodr\'{\i}guez-Ardila
          \inst{1}
          \and
          Miriani G. Pastoriza
          \inst{2}
          \and
          Sueli Viegas
          \inst{3}
          \and
          T.\ A.\ A.\ Sigut
          \inst{4}
          \and
          Anil K.\ Pradhan
          \inst{5}
          }

   \offprints{A. Rodr\'{\i}guez-Ardila}

   \institute{Laborat\'orio Nacional de Astrof\'{\i}sica $-$ Rua dos
Estados Unidos 154 $-$ Bairro das Na\c c\~oes. CEP 37500-000, Itajub\'a,
MG, Brazil.\\
              \email{aardila@lna.br}
         \and
         Departamento de Astronomia, Instituto de F\'{\i}sica,
UFRGS. Av. Bento Gon\c calves 9500, Porto Alegre, RS, Brazil. \\
              \email{mgp@if.ufrgs.br}
         \and
         Instituto de Astronomia, Geof\'{\i}sica e Ci\^encias
Atmosf\'ericas. Rua do Mat\~ao 1226. CEP 05508-900, S\~ao Paulo, SP,
Brazil.    \\
             \email{viegas@astro.iag.usp.br}
        \and
         Department of Physics and Astronomy,
         The University of Western Ontario, London, ON, Canada, N6A 3K7 \\
        \and
         Department of Astronomy, The Ohio State University, Columbus, OH, USA
         43210-1106  \\
        }

   \date{Received; accepted}

   \abstract{Near-infrared spectroscopy is used to study the kinematics and
excitation mechanisms of \h2\ and \fe2\ gas in a sample of mostly
Seyfert~1 galaxies. The spectral coverage allows simultaneous
observation of the JHK bands, thus eliminating the aperture and seeing
effects that have usually plagued previous works.  The \h2\ lines are
unresolved in all objects in which they were detected while the
\fe2\ lines have widths implying gas velocities of up to 650~\kms. This
suggests that, very likely, the \h2\ and \fe2\ emission does not
originate from the same parcel of gas.  Molecular \h2\ lines were
detected in 90\% of the sample, including PG objects, indicating  detectavel
amounts of molecular material even in objects with low levels of
circumnuclear starburst activity. Analysis of the observations favors
thermal excitation mechanisms for the \h2\ lines.
Indeed, in NGC\,3227, Mrk\,766, NGC\,4051 and NGC\,4151, the molecular
emission is found to be purely thermal but with heating processes which
vary among the objects.  Thermal excitation is also confirmed by the
rather similar vibrational and rotational temperatures in the objects
for which they were possible to derive. \fe2\ lines are detected in all
of the sample AGN. The \fe2\ 1.254$\mu$m/Pa$\beta$ ratio is compatible
with excitation of the \fe2\ lines by the active nucleus in most
Seyfert~1 galaxies, but in Mrk\,766 the ratio implies a stellar
origin.  A  correlation between \h2/Br$\gamma$ and \fe2/Pa$\beta$
is found for our sample objects supplemented with data from the
literature.  The correlation of these line ratios is a useful
diagnostic tool in the NIR to separate emitting line objects by their
level of nuclear activity. X-ray excitation models are able to explain
the observed \h2\ and part of the \fe2\ emission but fails to explain
the observations in Seyfert~2 galaxies.  Most likely, a combination of
X-ray heating, shocks driven by the radio jet, and circumnuclear star
formation contributes, in different proportions, to the \h2\ and
\fe2\ lines observed. In most of our sample objects, the
\fe2\ 1.257$\mu$m/1.644$\mu$m ratio is found to be 30\% lower than the
intrinsic value based on current atomic data.  This implies either than
the extinction towards the \fe2\ emitting clouds is very similar in
most objects or there are possible inaccuracies in the {\it A}-values
in the \ion{Fe}{ii} transitions.

   }

\maketitle

\section{Introduction}

One of the fundamental problems in active galactic nuclei (AGN)
research is to determine the dominant excitation mechanisms of the
narrow line emitting gas, i.e.\ whether it is due to non-stellar
processes (e.g.\ photoionization from a central source or shocks from a
radio jet) or to stellar processes (e.g.\ photoionization from OB stars
or shocks from supernova remnants).  This ambiguity is most evident
when interpreting the spectral lines of low-ionization species such as
\fe2\ and molecular hydrogen. Both sets of lines are detected in
galaxies displaying varying degrees of nuclear activity, from objects
classified as starburst-dominated to those classified as
AGN-dominated. Do these lines are produced by a unique mechanism
in the objects where they are observed, evidencing the
so-called AGN-starburst connection? or are they formed by different
mechanisms (or some combination thereof) in different classes of
objects?

Transitions within the ground state of \h2\ are often bright in
near-infrared observations of AGN.  The origin of these lines has been
highly debated, with no definitive consensus yet reached.  Three types
of excitation mechanisms may produce the \h2\ lines:  (a) UV
fluorescence (e.g Black \& Van Dishoek \cite{bvd}); (b) shocks
(Hollenbach \& McKee \cite{hm89}); and (c) X-ray illumination (Maloney
et al. \cite{mht96}). UV fluorescence is usually regarded as a
non-thermal excitation process and shocks and X-rays as thermal ones.
Theoretically, each of these processes produces a different spectrum,
and the relative intensities between emission lines of H$_{2}$ can be
used to discriminate the dominant excitation mechanism.  So far,
however, the results are controversial. For example, Veilleux et
al.\ (\cite{vsk97}), claim that shocks associated  with nuclear
outflows are the likely source of H$_{2}$ in AGN while Quillen et al.
(\cite{q99}) relate the H$_{2}$ emission to local star formation. Very
recently, Rigopoulou et al.  (\cite{ri02}) carried out a survey of
molecular hydrogen emission using a sample of starburst and Seyfert
galaxies observed by ISO. They conclude that the \h2\ emission in
Seyferts is most likely a combination of that from photodissociation
regions, shock-heated gas, and X-ray heated gas, thus favoring thermal
excitation over the non-thermal.  However, most of the above
conclusions have been obtained using a small number of objects with a
clear bias towards Seyfert 2.  In addition, the aperture used in the
ISO observations is usually large which allows the contamination of the
nuclear spectra by the host galaxy.

Similar problems, namely production mechanisms and location of the
emitting region, are associated with the interpretation of
\fe2\ emission in AGN.  Forbes \& Ward (\cite{fw93}) found a tight
correlation between [Fe\,{\sc ii}] and the 6-cm radio emission for both
starburst (SB) and Seyfert galaxies.  This was taken as strong evidence
of shock excitation from the radio jets and/or SNRs for the production
of [Fe\,{\sc ii}] in AGN. However, Simpson et al.\ (\cite{sfbw96})
argued that photoionization by the central engine must be the dominant
excitation mechanism. More recently, Mouri et al.\  (\cite{mkt00})
support this scenario by proposing that heating by X-rays is more
important in the production of \fe2\ and that photoionization by a
central source or by shocks can be distinguished by the electron
temperature of the \fe2\ region:  8000\,K in the former case and
6000\,K for the later. Another useful indicator proposed to distinguish
between SB and Seyfert activity for the origin of \fe2\ is the
ratio \fe2 1.257$\mu$m/Pa$\beta$ (Goodrich et al. \cite{gvh94};
Alonso-Herrero et al. \cite{alh97}). Observationally, there is an increasing
progression in that ratio from pure photoionization to pure shock
excitation. In SB galaxies it is observed to be in the range of
0.1--0.5 (Simpson et al. \cite{sfbw96}), in reasonable agreement with
the prediction of SB models of Colina (\cite{co93}), while in AGN it is
expected to be in the range of 0.6--3 if heating by X-rays dominates.

In order to address the above problems and to increase our knowledge
about the origin of the \h2\ and \fe2\ lines, we have observed a sample
of 19 Seyfert 1, 2 Seyfert 2 and 1 starburst galaxies. These objects
are part of an
ongoing program aimed at studying the near-infrared properties of AGN.
For most of our target objects, we provide the first reports of
molecular and \fe2\ emission. The remaining paper is organized in the
following manner: \S 2 presents the observations and data reduction; \S
3 discusses the kinematics of the molecular and \fe2\ gas inferred from
their line profiles; \S 4 discusses the origin of the \h2\ lines; \S 5
deals with the emission mechanisms leading to the \fe2\ lines;
concluding remarks are given in \S 6.

\section{Observations and data reduction}

Near-infrared (NIR) spectra from 0.8-2.4 $\mu$m were obtained at the
NASA 3\,m Infrared Telescope Facility (IRTF) on April 20-25/2002 with the
SpeX spectrometer (Rayner et al. \cite{ray03}). The detector consisted
of a 1024$\times$1024 ALADDIN 3 InSb array with a spatial scale of
0.15''/pixel. Simultaneous wavelength coverage was obtained by means of
prism cross-dispersers.  A 0.8''$\times$15'' slit was used during the
observations, giving a spectral resolution of 360\kms. For the objects
in which the host galaxy is clearly detected on the DSS images, the
slit was oriented along the major axis of the target. Otherwise, it was
aligned to the parallactic angle in order to minimize slit loses
because of differential refraction. During the different nights, the
seeing varied between 0.7''--1''. Observations were done nodding in an
ABBA source pattern with typical individual integration times of 120\,s
and total on-source integration times between 30 and 50 minutes.  For
some sources, we took spectra on subsequent nights which were, after
reduction, combined to form a single spectrum. During the observations,
A0\,V stars were observed near each target to provide telluric
standards at similar air masses. They were also used to flux calibrate
the sample.  Table~\ref{log} shows the log of the observations and the
atmospheric conditions during each night. The galaxies are listed in
order of right ascension.

The spectral extraction and wavelength calibration procedures were
performed using SPEXTOOL, the in-house software developed and provided
by the SpeX team for the IRTF community (Cushing et al. \cite{cvr03};
Vacca et al. \cite{vcr03}) \footnote{SPEXTOOL is available from the
IRTF web site at http://irtf.ifa.hawaii.edu/Facility/spex/spex.html}.
Each pair of AB observations was reduced individually and the results
averaged to provide a final spectrum.  A 1.6''$-$2'' aperture, depending on
the seeing conditions during the observations, was used to extract the
spectra.  Observations of an argon arc lamp enabled wavelength
calibration of the data; the rms of the dispersion solution was, on
average, 0.18 \AA. No effort was made to extract spectra at positions
different from the nuclear region even though some objects show
evidence of extended emission. The small beam employed (usually
1.8''$\times$0.8'') and the good seeing during the observations makes
this data set truly representative of the nuclear emission, with little
contamination of the host galaxy or the extended narrow line region.

\begin{table*}
\begin{center}
\caption{Observation log and basic galactic properties for the
sample. The objects are listed in order of right ascension.} \label{log}
\begin{tabular}{llcccccccc}
\hline \hline
   &         &      &     &        &             & On-source  &     &    & \\
   &         &      &     &        & Date of     & Integration &     & PA & r$^{\mathrm{a}}$ \\
ID & Galaxy  & Type & $z$ & E(B-V)$_{\mathrm{G}}$ & Observation & time  (s)   & Airmass & (\degr) & (pc) \\
(1) & (2) & (3) & (4) & (5) & (6) & (7) & (8) & (9) & (10) \\
\hline
1  & Mrk\,1210 & Sy2  & 0.01406 & 0.030 & 2002 Apr 25 & 2700 & 1.25 & 58 & 220 \\
2  & Mrk\,124  & NLS1 & 0.05710 & 0.015 & 2002 Apr 23 & 2640 & 1.16 & 10 & 990 \\
3  & Mrk\,1239 & NLS1 & 0.01927 & 0.065 & 2002 Apr 21 & 1920 & 1.08 & 0.0 & 335 \\
   &           &      &         &       & 2002 Apr 23 & 1920 & 1.15 & 0.0 &  \\
4  & NGC\,3227 & Sy1  & 0.00386 & 0.023 & 2002 Apr 21 & 720  & 1.00 & 158 & 67 \\
   &           &      &         &       & 2002 Apr 25 & 1080 & 1.02 & 158 &  \\
5  & H1143-192 & Sy1  & 0.03330 & 0.039 & 2002 Apr 21 & 1920 & 1.31 & 45  & 520 \\
6  & NGC\,3310 & SB   & 0.00357 & 0.022 & 2002 Apr 21 & 840  & 1.21 & 158 & 56 \\
7  & NGC\,4051 & NLS1 & 0.00234 & 0.013 & 2002 Apr 20 & 1560 & 1.17 & 132 & 37 \\
8  & NGC\,4151 & Sy1  & 0.00345 & 0.028 & 2002 Apr 23 & 1800 & 1.10 & 130 & 58 \\
9  & Mrk\,766  & NLS1 & 0.01330 & 0.020 & 2002 Apr 21 & 1680 & 1.06 & 112 & 230 \\
   &           &      &         &       & 2002 Apr 25 & 1080 & 1.02 & 112 & \\
10 & NGC\,4748 & NLS1 & 0.01417 & 0.052 & 2002 Apr 21 & 1680 & 1.29 & 36 & 254 \\
   &           &      &         &       & 2002 Apr 25 & 1440 & 1.21 & 36 & \\
11 & Mrk\,279  & NLS1 & 0.03068 & 0.016 & 2002 Apr 24 & 3600 & 1.54 & 0.0 & 480 \\
12 & NGC\,5548 & Sy1  & 0.01717 & 0.020 & 2002 Apr 23 & 1920 & 1.05 & 112 & 298 \\
13 & Mrk\,478  & NLS1 & 0.07760 & 0.014 & 2002 Apr 20 & 3240 & 1.06 & 0.0 & 1200 \\
14 & NGC\,5728 & Sy2  & 0.01003 & 0.101 & 2002 Apr 21 & 960  & 1.31 & 36 & 160 \\
15 & PG\,1448+273 & QSO & 0.06522 & 0.029 & 2002 Apr 24 & 2160 & 1.01 & 108 & 1020 \\
16 & Mrk\,291  & NLS1 & 0.03519 & 0.038 & 2002 Apr 21 & 2520 & 1.04 & 84 & 550 \\
17 & Mrk\,493  & NLS1 & 0.03183 & 0.025 & 2002 Apr 20 & 1800 & 1.07 & 0.0 & 500 \\
   &           &      &         &       & 2002 Apr 25 & 900  & 1.04 & 0.0 & \\
18 & PG\,1612+261 & QSO & 0.13096 & 0.054 & 2002 Apr 23 & 2520 & 1.10 & 107 & 2050 \\
19 & Mrk\,504  & NLS1 & 0.03629 & 0.050 & 2002 Apr 21 & 2100 & 1.04 & 138 & 570 \\
20 & 1H\,1934-063 & NLS1 & 0.01059 & 0.293 & 2002 Apr 21 & 1200 & 1.13 & 150 & 170\\
   &           &      &         &       & 2002 Apr 25 & 720  & 1.17 & 150  & \\
21 & Mrk\,896  & NLS1 & 0.02678 & 0.045 & 2002 Apr 23 & 1440 & 1.21 & 150 & 420 \\
   &           &      &         &       & 2002 Apr 24 & 1200 & 1.18 & 150 & \\
   &           &      &         &       & 2002 Apr 25 & 1200 & 1.17 & 150 & \\
22 & Ark\,564  & NLS1 & 0.02468 & 0.060 & 2000 Oct 10 & 1500 & 1.05 & 0.0 & 390 \\
\hline
\end{tabular}
\end{center}
$^{\mathrm{a}}$ Radius of the integrated region.
\end{table*}

Final wavelength-calibrated target spectra were divided by the
spectrum of an A0V star observed at a similar airmass to eliminate
telluric contamination. Typical airmass differences between the target
and the standard were 0.05-0.1. However, despite these small
differences, a small residual, mainly in the interval
1.995$\mu$m$-$2.087$\mu$m, was found in some objects due to bad telluric band
cancellation.  After division by the A0V standard, each spectrum was
multiplied by a blackbody of temperature equal to the effective
temperature of the star to restore the true continuum
shape of the target.

\begin{figure*}
   \centering
   \includegraphics[width=14cm]{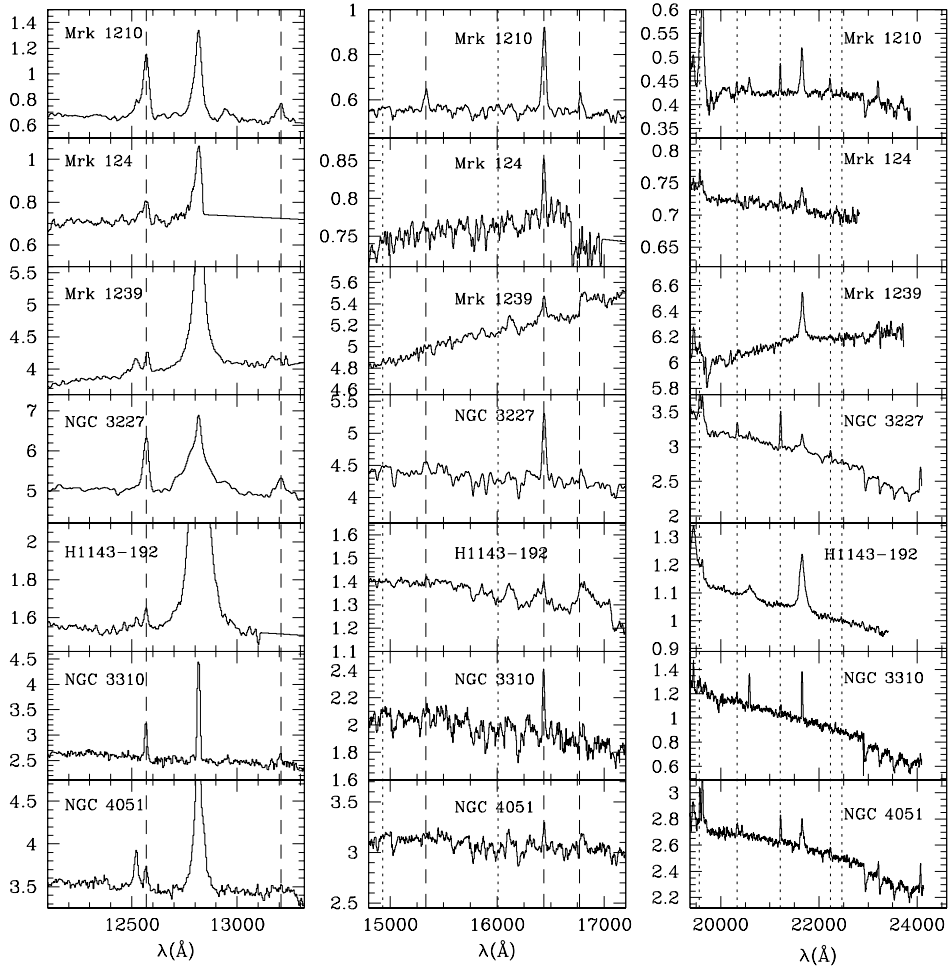}
      \caption{Final reduced spectra, in the Earth's frame, centred in
near Pa$\beta$ (left panel), the H~band (1.60$\mu$m, middle panel),
and Br$\gamma$ (right panel). The abscissa is the observed flux in
units of 10$^{-15}$ erg cm$^{-2}$ s$^{-1}$ \AA$^{-1}$. The identified
\fe2\ (dashed lines) and \protect\h2\ lines (dotted lines) are marked in the
spectra. Note that in the H band, no \h2\ lines are detected except
in NGC\,5728 and NGC\,3227 where (1-0)S(7) 1.747$\mu$m is visible.
We show, however, the expected position of two of the strongest lines
predicted by the Black \& van Dishoeck's (1987) models.
              }
         \label{fig1a}
   \end{figure*}

\begin{figure*}
   \centering
   \includegraphics[width=14cm]{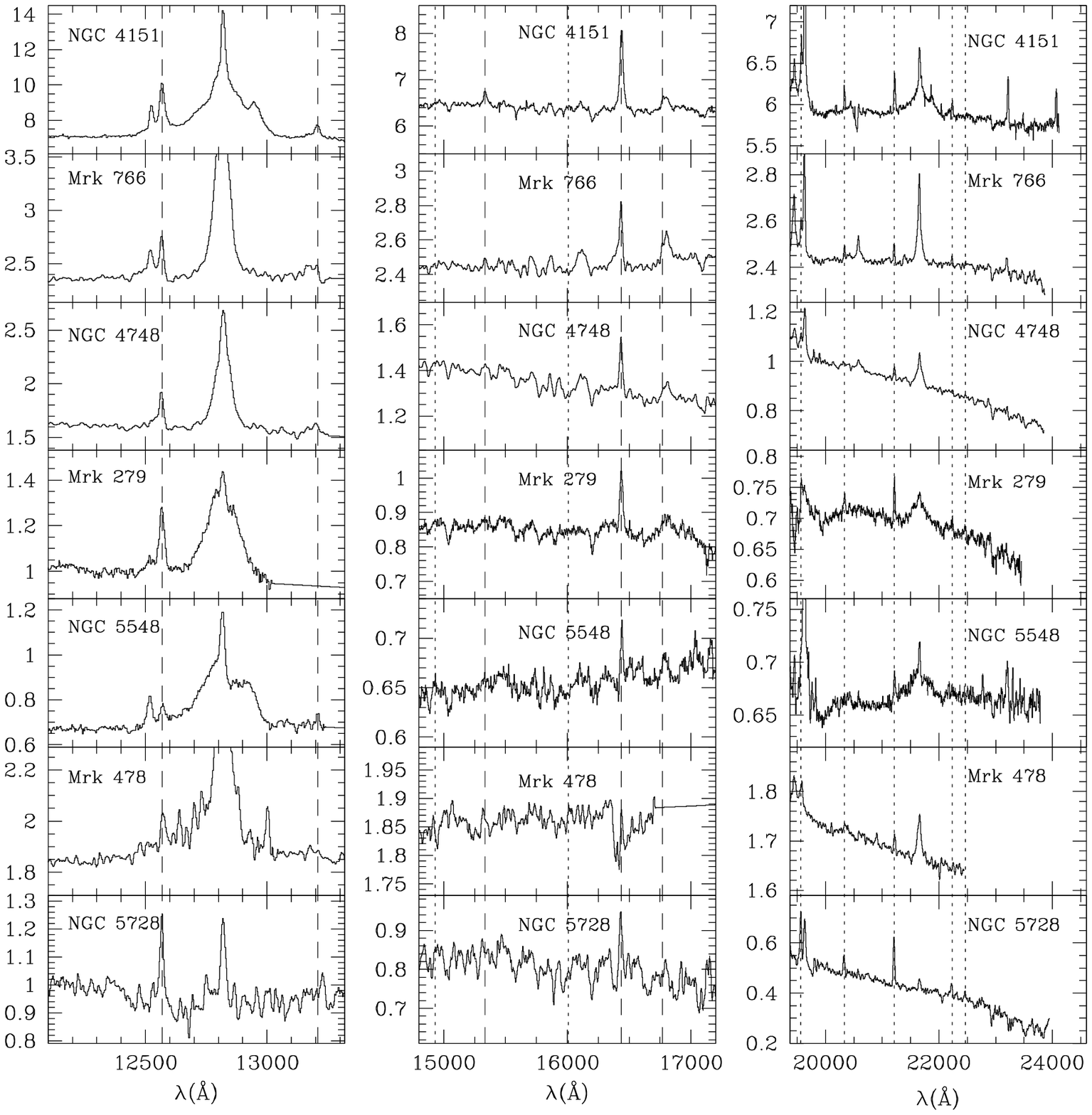}
      \caption{Same as Fig.~\ref{fig1a}.
              }
         \label{fig1b}
   \end{figure*}

\begin{figure*}
   \centering
   \includegraphics[width=14cm]{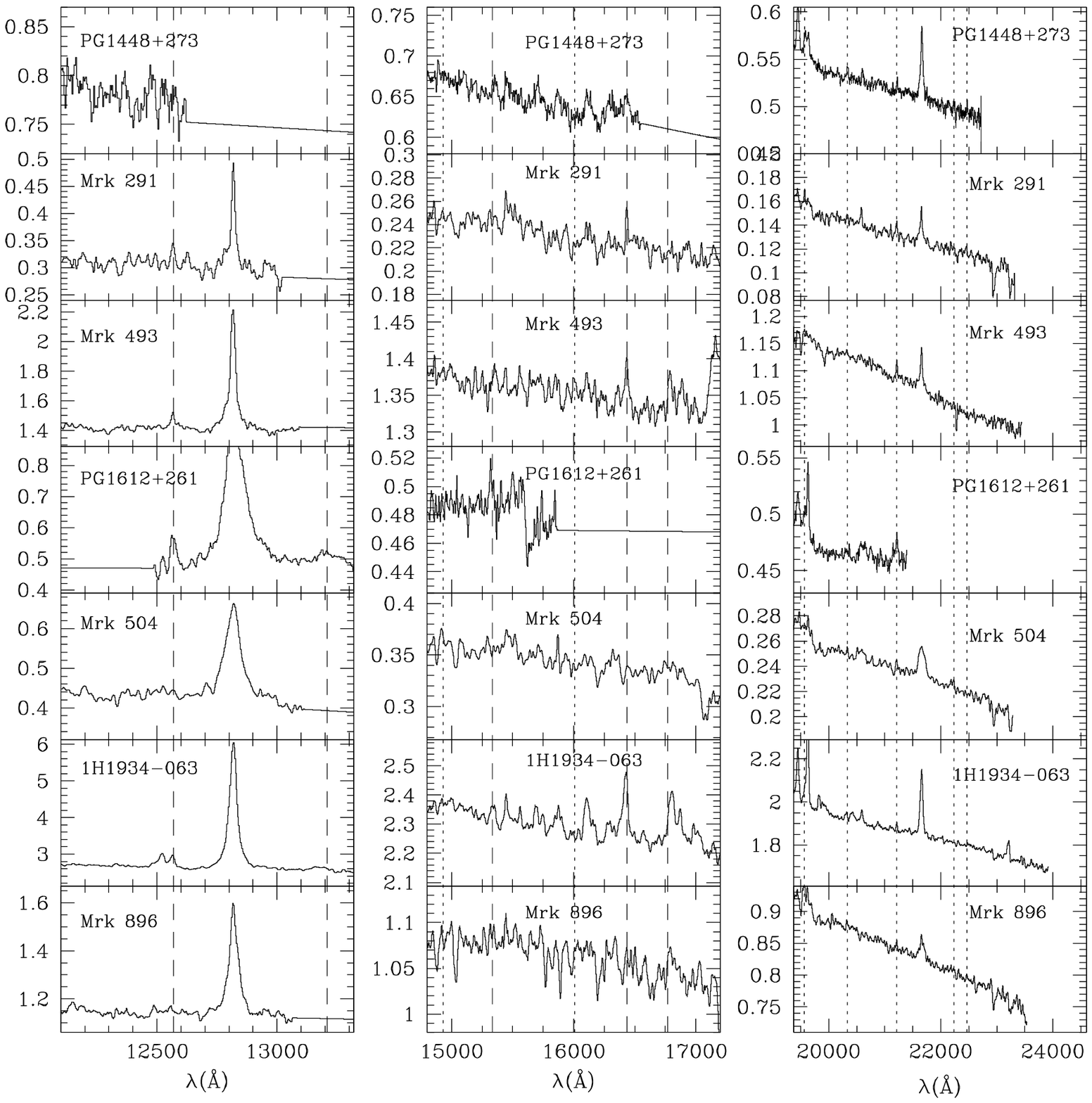}
      \caption{Same as Fig.~\ref{fig1a}.
              }
         \label{fig1c}
   \end{figure*}

Flux calibration was carried out by normalizing to the K-band magnitude
of the corresponding telluric standard spectrum.  We estimate that the
uncertainty in this procedure is $\sim$10\% by comparing our
calibration with the 2MASS fluxes available in the literature. The
agreement in the continuum flux level in the overlap region for the
different dispersion orders was excellent with errors less than 1\%.
The spectra were then corrected for redshift, determined from the
average $<z>$ measured from the positions of [S\,{\sc iii}]
0.953$\mu$m, Pa$\delta$, He\,{\sc i} 1.083$\mu$m, \ion{O}{i}
1.128$\mu$m, Pa$\beta$ and Br$\gamma$.  Finally, a Galactic extinction
correction as determined from the {\it COBE/IRAS} infrared maps of
Schlegel, Finkbeiner \& Davis (1998) was applied. The value of the
Galactic {\it E(B-V)} used for each galaxy is listed in Col.~5
of Table~\ref{log}. Final reduced spectra, in the spectral regions of
interest in this work, are plotted in Figures~\ref{fig1a}
to~\ref{fig1c}.

\section{Kinematics of the H$_{2}$ and \fe2\ lines} \label{kin}

The large spectral coverage and medium spectral resolution of our data
(360 km s$^{-1}$) allow us to discuss how the widths of the H$_{2}$
lines compares to that of other narrow lines. The goal of this analysis
is to check if the molecular hydrogen is kinematically linked to the
NLR gas, setting important constrains on its location and origin. To
our knowledge, no previous results about this issue are available in
the literature, mainly because of the lack of adequate spectral
resolution in the NIR. Recent NICMOS imaging of  \h2\ emission
by Quillen et al.\ (\cite{q99}) for a sample of Seyfert galaxies found
H$_{2}$ on scales of a few hundred parsecs from the nucleus for three
objects (NGC\,5643, NGC\,2110 and Mrk\,1066).  The \h2\ emission was
found to be coincident with [O\,{\sc iii}] and H$\alpha$+[N\,{\sc ii}],
suggesting that the molecular gas may follow the NLR gas.
We can test this connection by examining the the line profiles:  if the
line widths of the molecular gas are similar to that of other narrow
emission lines, then, under the assumption that the widths of the NLR
lines reflect the large scale motions of the emitting clouds in the
gravitational potential of the central mass, similar widths of
molecular and forbidden lines suggest that they are co-spatial.

Table~\ref{FWHM} lists the FWHM in \kms\ measured in [\ion{S}{iii}]
0.9531$\;\mu$m, [\ion{Si}{vi}] 1.963$\;\mu$m, [Fe\,{\sc ii}]
1.2567$\;\mu$m and H$_{2}$ (1,0)S(1) 2.121$\;\mu$m. These values
were obtained by fitting a Gaussian to the observed profile and then
adopting the FHWM of the fit as representative of the FWHM of the line.
The software LINER (Pogge \& Owen \cite{pow93}) was used for this
purpose. Afterwards, we corrected the measurements for instrumental
resolution. In all cases, the lines in Table~\ref{FWHM} were
well represented by a single Gaussian component. For [\ion{S}{iii}],
[Fe\,{\sc ii}] and H$_{2}$ (1,0)S(1) 2.121$\mu$m, no
statistically significant shifts relative to the fiducial {\it z}
were found while the centroid of [\ion{Si}{vi}] was observed to be
blueshifted by up to a few hundreds \kms\ in some objects,
as is typical for coronal lines (Penston et al. \cite{pe84}).
Because this latter line is out of the main focus of this paper,
we will not make further comments about its shifts.
Typical errors, based on the uncertainties in
the placement of the continuum, are $\sim$30 km~s$^{-1}$. We have
chosen H$_{2}$~(1,0)S(1)~2.121$\mu$m as representative of molecular
hydrogen because it is one of the strongest \h2\ line in the NIR.
In addition, it is isolated from other emission features, minimizing
the effects of blending. Note that the the width and shape of other \h2\
lines within a given galaxy is rather similar to that of
H$_{2}$~2.121$\mu$m. No \ion{H}{i} lines were used in this analysis
because, for more than half of the sample, the deblending of the
observed profile into the components emitted by the NLR and BLR is
highly uncertain.

The results show that in all objects the width of H$_{2}$ (1,0)S(1) is
at the limit of the spectral resolution (360 km s$^{-1}$) while
\fe2\ is resolved in 50\% of the sample.  [S\,{\sc iii}] 0.953 $\mu$m
(IP=23.3 eV), considered the NIR counterpart of [O\,{\sc iii}] 5007
\AA, is unresolved in 60\% of the sample, and in the remaining 40\%,
the trend is to have similar or larger width than [Fe\,{\sc ii}]. The coronal line
[Si\,{\sc vi}] 1.963$\mu$m (IP=167 eV), when resolved, is significantly
broader than H$_{2}$ (1,0)S(1) and, except for a few cases, is also
broader than [Fe\,{\sc ii}]. Overall, in those objects in which the
forbidden lines were resolved, we found increasing line widths with
increasing ionization potential of the emitting species. A similar
trend using optical and NIR lines was found by Rodr\'{\i}guez-Ardila et
al. (\cite{ro02}). They showed, for a small sample of Seyfert~1
galaxies, a positive correlation between FWHM and ionization potential.
In the optical region, this type of correlation has been firmly
established for a large number of AGN (Penston et al. \cite{pe84}; De
Robertis \& Osterbrock \cite{roo84}, Evans \cite{ev88}). This trend is
usually interpreted as a stratification of the emitting regions in the
sense that the highest ionized gas is located closer to the central
source than low-ionization/neutral NLR gas.

NGC\,3227, NGC\,4151 and PG\,1612+261 are counter-examples to the above
trend. In these objects, [\ion{Fe}{ii}] is the broadest NLR line, even
when compared to high ionization lines such as [\ion{S}{ix}]
1.252$\mu$m (IP=328 eV).  From a kinematical point of view, the bulk of
the [\ion{Fe}{ii}] and \h2\ emission cannot originate in the same
volume of gas. Part of [\ion{Fe}{ii}] must arise from an additional
source. We suggest that this extra source may be associated with shock
excitation from the radio jet.  In fact, Knop et al.\ (\cite{knop96})
proposed this scenario for NGC\,4151. They found [\ion{Fe}{ii}] broader
than the narrow component of Pa$\beta$ and [\ion{S}{ix}], the latter
two lines also unresolved in their nuclear spectrum. Knop's FWHM for
[\ion{Fe}{ii}] in NGC\,4151 is close to ours (434\kms\ vs 505\kms,
respectively).

\begin{table*}
\begin{center}
\caption{FWHM (in km s$^{-1}$), corrected for instrumental resolution,
for H$_{2}$ and other important lines measured in spectra sample.} \label{FWHM}
\begin{tabular}{lccccc}
\hline \hline
      & [\ion{S}{iii}] & [\ion{Si}{vi}] & \fe2\ & \h2\ (1,0)S(1) & \ion{O}{i} \\
Galaxy &  0.9530$\mu$m &  1.963$\mu$m  &  1.257$\mu$m    & 2.121$\mu$m  & 1.128$\mu$m \\
\hline
  Mrk\,1210     &  685 & 660  &    650   &   380   & ... \\
  Mrk\,124      &  420 & 360  &    420   &   395   & 1330  \\
  Mrk\,1239     &  865 & ...  &    456   &   ...   & 1160 \\
  NGC\,3227     &  470 & 420  &    560   &   360   & 3110 \\
  H\,1143-192   &  360 & 556  &    404   &   ...   & 1630 \\
  NGC\,3310     &  360 & ...  &    360   &   360   &  ... \\
  NGC\,4051     &  360 & 360  &    360   &   360   & 880 \\
  NGC\,4151     &  360 & 360  &    505   &   370   & 3520 \\
  Mrk\,766      &  360 & 390  &    370   &   370   & 1270 \\
  NGC\,4748     &  360 & 590  &    440   &   370   & 1550 \\
  Mrk\,279      &  610 & ...  &    440   &   370   & 2790 \\
  NGC\,5548     &  360 & 480  &    435   &   360   & 4860 \\
  Mrk\,478      &  360 & ...  &    370   &   380   & 1160 \\
  NGC\,5728     &  405 & 520  &    360   &   360   & ... \\
  PG\,1448+273  &  370 & 526  &    ...   &   370   & 728 \\
  Mrk\,291      &  360 & ...  &    370   &   370   & ... \\
  Mrk\,493      &  360 & ...  &    380   &   382   & 610 \\
  PG\,1612+261  &  360 & 450  &    550   &   370   & 1900 \\
  Mrk\,504      &  617 & 430  &    ...   &   360   & 1630 \\
  1H\,1934-063  &  420 & 650  &    550   &   360   & 850 \\
  Mrk\,896      &  360 & ...  &    360   &   360   & 1050 \\
  Ark\,564      &  440 & 530  &    360   &   370   & 600  \\

\hline
\end{tabular}
\end{center}
\end{table*}

The smaller values of the FWHM of H$_{2}$ relative to other narrow
lines may suggest that the molecular gas is concentrated in the
external/extended NLR or possibly even in the host galaxy,
where the gravitational effects of the black hole on the emitting
gas are lower than in the BLR. This scenario seems unlikely, however,
because for the large majority of the objects the size of the integrated
region covered by the spectra is no larger than 500~pc (see
Col.~10 of Table~\ref{log}). The alternative is to consider
that the molecular gas is not kinematically
linked to the NLR gas even though the two are co-spatial.

\begin{figure}
   \centering
   \includegraphics[width=8cm]{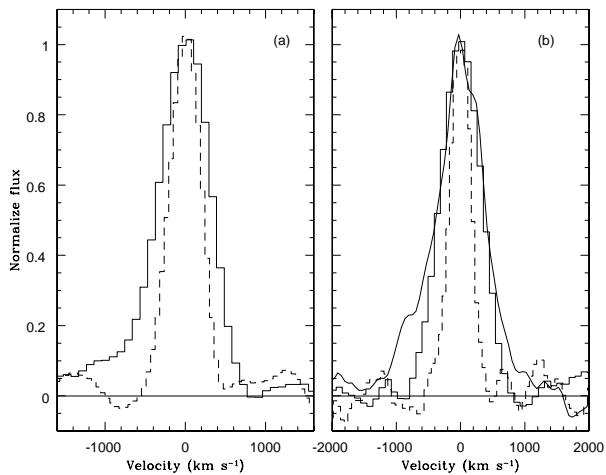}
   \caption{Comparison of the line profiles of [\ion{Fe}{ii}] 1.257$\mu$m
(full histogram) and \h2\ 2.122$\mu$m (dashed histogram) for (a) NGC\,3227
and (b) Mrk1210, for which the observed profile of Br$\gamma$ (solid line)
has also been plotted.
          }
              \label{perfis}
    \end{figure}

NGC\,3227 supports this scenario.
Figure~\ref{perfis}a shows a comparison of the line profiles of
\h2~2.121$\mu$m and [\ion{Fe}{ii}]~1.257$\mu$m. It can be seen that the
latter line is much broader than the former although both
are emitted from a region $\sim$70~pc in radius. Note that
[\ion{Fe}{ii}]~1.257$\mu$m is slightly asymmetric
towards the blue, mainly in the wing of the line, but it is not
possible to tell from the data if the asymmetry is related to
\fe2\ or it is due to \ion{He}{i} 1.252$\mu$m. NICMOS imaging
in \h2~(1-0)S(1)~2.121$\mu$m
carried out by Quillen et al. (\cite{q99}) on this object shows that
the molecular hydrogen emission originates in a 100\,pc diameter disk
that is not directly associated with either the [O\,{\sc iii}] emission
or the 18\,cm radio emission. Moreover, Rodr\'{\i}guez-Ardila \& Viegas
(\cite{rv03}) detected 3.3$\mu$m PAH emission in the inner 120\,pc of
this object, implying circumnuclear star formation within that region
probably related to the disk of molecular gas. The differences in line
widths reported here supports the view that both the NLR and the
molecular gas are co-spatial but emitted by different volumes of gas
and most probably originating from different excitation mechanisms.

Mrk\,1210 is another interesting case for a kinematical decoupling of
the molecular gas. The forbidden NLR lines are remarkably similar in
FWHM (see Table~\ref{FWHM}); however, H$_{2}$ (1-0)S(1) 2.121$\mu$m is
significantly narrower, almost half the FWHM value of the [Fe\,{\sc
ii}] 1.257$\mu$m (Figure~\ref{perfis}b). Radio observations in the
CO(1$\rightarrow$0) 115 GHz and CO(2$\rightarrow$1) 230 GHz carried out
on this object (Raluy et al. \cite{rpc98}) show a high level of star
forming activity within the inner 5\,Kpc. Moreover, this Seyfert 2 has
a strong H$_{2}$0 megamaser of nuclear origin, arising from the
innermost parsecs of the host galaxy (Braatz et al. \cite{bra94}).
Because our integrated spectrum covers a region of $\sim$200~pc in
radius, it is
likely that it includes a significant contribution from the
circumnuclear starburst, and the \h2\ emission we observe can be
directly associated to this component.

NGC\,5728 also provides additional support to the idea of disconnection
between the \h2\ emission and the NLR.  This AGN is a classic
example of a Seyfert~2 galaxy with a biconical emission line region
separated by a dark band; the nucleus lies behind, at the point defined
by the apexes of the two ionization cones (Wilson et al. \cite{wil93}).
The ionization cones are essentially collinear with the position angle
of the brighter SE cone being 118\degr. In addition to the cones, the
nuclear region of NGC\,5728 shows other interesting features. In red
and green optical continuum images presented by Wilson et al.
(\cite{wil93}), there are two miniature bars delineated by four main
knots of emission, making the central region of NGC\,5728 very complex
with several decoupled dynamical components. In its K-band spectrum,
NGC\,5728 shows a remarkable \h2\ emission; it is the only object in
which the \h2\ emission is stronger than Br$\gamma$.  Our spectrum
covers the inner 320\,pc with the slit oriented along the larger bar
(PA=36$^{\rm o}$), almost perpendicular to the ionization cones and
along the dark band. This suggest that the \h2\ emission is associated
with the dark band, which serves as a natural reservoir of molecular
gas.

We also note that Mrk\,1239 and H\,1143-192 are the only Seyfert 1
galaxies of our sample with no \h2\ lines observed in their spectra. As
\fe2\ emission and other NLR features are clearly visible in their
spectra (see Fig.~\ref{fig1a}), these objects provide additional
support for a separate spatial origin between the \h2\ and NLR lines.

Finally from Table~\ref{FWHM}, the molecular and forbidden lines are
all significantly narrower than the BLR lines. The last column of
Table~\ref{FWHM} shows the FWHM of \ion{O}{i} 1.128$\mu$m, a pure BLR
feature emitted in the outer portions of that region. The large
difference in line widths between \h2\ and \ion{O}{i} makes implausible
the emission of \h2\ close to the BLR, i.e.\ from the torus of gas and
dust around the central engine.

\subsection{NIR kinematics vs optical kinematics}

It is interesting to see how the widths of NIR lines
measured in \S~\ref{kin} for the galaxy sample
compares to those found in the optical region.
This will allow us to get an insight of
how the NLR gas is distributed and most importantly,
to check if large amounts of high-velocity gas are
obscured by dust extinction at shorter wavelengths.
For this analysis we have chosen the lines
[\ion{O}{iii}] $\lambda$5007 and [\ion{S}{ii}] $\lambda$6717.
The former can be considered as the optical counterpart
of [\ion{S}{iii}] $\lambda$9531 while the latter is
representative of low-ionization gas as [\ion{Fe}{ii}] is.
Values of FWHM for these two lines, extracted from the
literature, are listed in Table~\ref{fwhmlit}. Note that
for most cases, the spectral resolution is higher than that of
NIR lines.

\begin{table}
\begin{center}
\caption{Values of FWHM, in km s$^{-1}$, for optical lines of the
galaxy sample taken from the literature} \label{fwhmlit}
\begin{tabular}{lcc}
\hline\hline
Galaxy    &   [\ion{O}{iii}] $\lambda$5007  & [\ion{S}{ii}] $\lambda$6717 \\
\hline
Mrk\,1210  &     494$^{2}$   &     276$^{2}$ \\
Mrk\,1239  &     630$^{1}$   &     360$^{a,3}$ \\
NGC\,3227  &     460$^{5}$   &     427$^{5}$ \\
NGC\,4051  &     162$^{5}$   &     185$^{5}$ \\
NGC\,4151  &     245$^{5}$   &     280$^{2}$ \\
Mrk\,766   &     180$^{5}$   &     159$^{5}$  \\
NGC\,4748  &     365$^{1}$   &     450$^{a,3}$ \\
NGC\,5548  &     411$^{6}$   &     300$^{6}$ \\
Mrk\,478   &     920$^{1}$   &     ... \\
PG\,1448+273 &   315$^{1}$   &     ... \\
Mrk\,493   &     450$^{1}$   &     ... \\
1H\,1934-063 &   562$^{3}$   &     446$^{a,3}$ \\
Mrk\,896   &     315$^{1}$   &     ... \\
Ark\,564   &     265$^{4}$   &     260$^{4}$ \\
\hline
\end{tabular}
\end{center}
$^{a}$ FWHM of [\ion{N}{ii}] $\lambda$6584. \\

References $-$ 1. Veron-Cetty et al. 2001; 2. Schulz \& Henkel 2003;
3. Rodr\'{\i}guez-Ardila et al. 2000; 4. Erkens et al. 1997;
5. Veilleux, 1991; 6. Moore et al. 1996.
\end{table}

Overall, Table~\ref{fwhmlit} supports the conclusions drawn
in the previous section. Moreover, no significant broadening
of NIR lines relative to the optical ones is detected. The exceptions
are Mrk\,1210 and Mrk\,1239. These two sources are known to
have strong polarization due to dust scattering (Tran 1995;
Goodrich 1989), supporting the view that high velocity gas
are obscured by dust extinction. Note that for Mrk\,766 and
NGC\,4051, the optical lines are half as broad as the NIR ones,
but in both cases, the NLR lines are spectroscopically
unresolved.

It is worth commenting the results found for
NGC\,3227, NGC\,4151, NGC\,5548 and 1H\,1934-063. In
these objects, the width of [\ion{O}{iii}] is compatible
to that found for [\ion{S}{iii}], which argues against the 
presence of high-velocity gas in the visible region, not 
detected due to extinction. However, [\ion{Fe}{ii}]
is considerably broader than [\ion{S}{ii}] (or [\ion{N}{ii}]
where the measurements of the sulfur line were not
available). This gives additional support to the view that the
[\ion{Fe}{ii}] lines are at least partially formed in
a distinct region from that originating other low-ionization species.
Interestingly, the former three galaxies display
radio-jets, whose interaction with the NLR has been
proposed as the source of [\ion{Fe}{ii}] lines.

\section{Excitation mechanisms of the NIR H$_{2}$ lines} \label{sec4}

The \h2\ molecule can be excited via three distinct mechanisms:  (1)
{\it UV fluorescence}, where photons with $\lambda \leq$ 912 \AA\ are
absorbed by the \h2\ molecule and then re-emitted resulting in the
population of the various vibration-rotational levels of the ground
electronic state (e.g.\ see Black \& van Dishoek \cite{bvd}); (2) {\it
shocks}, where high-velocity gas motions heat, chemically alter, and
accelerate the ambient gas resulting in excitation of the \h2\ molecule
(e.g.\ see  Hollenbach \& McKee \cite{hm89}); and (3) {\it X-ray
illumination}, where hard X-ray photons penetrate deep into molecular
clouds, heating large amounts of molecular gas resulting in
\h2\ emission (e.g.\ see Maloney et al.  \cite{mht96}). Each of these three
mechanisms produces a different \h2\ spectrum; therefore, the relative
intensities between emission lines of H$_{2}$ can help to discriminate
the dominant excitation process.

Fluxes of the detected H$_{2}$ and permitted \ion{H}{i} lines are
listed in Table~\ref{h2flux}. They were measured by fitting a Gaussian to
the observed profile and then integrating the flux under the curve. The
LINER software (Pogge \& Owen \cite{pow93}) was used for this
purpose. For a large majority of our sample (16 out of 22
galaxies), these measurements are the first ones reported in the
literature.  H$_{2}$(1,0)S(1) was detected at a 3$\sigma$ level in 20
out of 22 galaxies, the exceptions being Mrk\,1239 and H\,1143-182 (see
Figs.~\ref{fig1a} to~\ref{fig1c}). It indicates that molecular
gas within the inner 500~pc is common in AGN and not restricted
to SB-AGN composite objects.  In addition, we clearly detected
H$_{2}$(1,0)S(2) for the large majority of AGN. We also detected
strong H$_{2}$(1,0)S(3) emission, sometimes larger than H$_{2}$(1,0)
S(1), but its flux must be used with caution because it may be severely
affected by telluric absorption.  Moreover, it is strongly blended with
Br$\delta$ to the blue and [Si\,{\sc vi}]~1.963$\mu$m to the red.
Fluxes for H$_{2}$(1,0)S(0) and H$_{2}$ (2,1)S(1) could also be
measured at the 3$\sigma$ level in 9 objects; fluxes were derived for
the remaining objects using the 3 $\sigma$ error as an upper limit.
In some Seyfert~1, it was possible to deblend the contribution of the
NLR from the observed Br$\gamma$ and Pa$\beta$ lines. Their fluxes
appears in Cols.~2 and~3, respectively, of Table~\ref{narrow}.

\begin{table*}
\caption[]{Observed fluxes of \h2, \fe2\ and \ion{H}{i} lines in units of 10$^{-15}$ erg~cm$^{-2}$~s$^{-1}$.} \label{h2flux}
\tiny{
\begin{tabular}{lccccccccccc}
\hline \hline
 &   [\ion{Fe}{ii}]  & Pa$\beta^{\mathrm{b}}$ & [\ion{Fe}{ii}] & \h2\ & \h2\  & \h2\  &  Br$\gamma^{\mathrm{b}}$ & \h2\  &  \h2 & 1.257$\mu$m/\\
Galaxy & 1.257$\mu$m & 1.282 $\mu$m & 1.644 $\mu$m & 1.957$\mu$m & 2.0332$\mu$m & 2.1213$\mu$m & 2.165 $\mu$m & 2.247$\mu$m & 2.223 $\mu$m & 1.644$\mu$m \\
\hline
Mrk\,1210  & 15.80$\pm$0.70 & 31.45$\pm$0.88 & 17.10$\pm$0.64 & 3.01$\pm$0.30 & 0.80$\pm$0.17 &  2.00$\pm$0.12 & 6.01$\pm$0.37 & 0.44$\pm$0.09 & 0.90$\pm$0.20 & 0.92$\pm$0.05  \\
Mrk\,124   & 3.16$\pm$0.62 & 14.6$^{a}$ & 2.82$\pm$0.47 & 0.91$\pm$0.16 & 0.36$\pm$0.05 &  0.90$\pm$0.12 & 2.49$\pm$0.55  &  $<$0.18  & 0.28$\pm$0.08 & 1.12$\pm$0.29 \\
Mrk\,1239  & 7.90$\pm$0.55 & 198.6$\pm$2.60 & 9.72$\pm$1.55 & $<$2.09 & $<$0.93  &  $<$0.58 &  32.0$\pm$1.50  & $<$0.96     & $<$0.82 & 0.81$\pm$0.14   \\
NGC\,3227  & 41.1$\pm$1.18 & 185.3$\pm$5.70 & 41.0$\pm$3.60 &  20.8$\pm$1.2  & 7.97$\pm$0.92 &  17.7$\pm$1.00 & 30.0$\pm$1.56 &  2.32$\pm$0.21 & 3.67$\pm$0.66 & 1.00$\pm$0.09 \\
H\,1143-192& 2.54$\pm$0.45 & 182.0$\pm$2.77 & 2.45$\pm$0.53 & $<$0.30  & $<$0.23 &  $<$0.12 & 26.7$\pm$1.00 & 0.24$\pm$0.08 & $<$0.14 & 1.04$\pm$0.28 \\
NGC\,3310  & 10.70$\pm$0.87 & 31.74$\pm$1.03 & 11.12$\pm$1.57 & $<$1.36 & 1.24$\pm$0.25 &  1.54$\pm$0.32& 10.79$\pm$0.33  & $<$0.32 & $<$0.67 & 0.91$\pm$0.15 \\
NGC\,4051  & 5.26$\pm$0.81 & 88.4$\pm$1.80 & 6.42$\pm$0.97 &  7.51$\pm$0.62 & 2.82$\pm$0.70 &  5.81$\pm$0.37 & 13.1$\pm$0.84 &  $<$0.56  & 1.52$\pm$0.40 & 0.82$\pm$0.18 \\
NGC\,4151  &  58.8$\pm$2.94 & 825.6$\pm$11.8 & 56.2$\pm$1.96 &  21.2$\pm$2.0  & 6.84$\pm$0.42 &  14.7$\pm$0.55 & 153.8$\pm$7.7 &  $<$0.62  & 5.04$\pm$0.62 & 1.04$\pm$0.06  \\
Mrk\,766   & 7.69$\pm$0.70 & 145.0$\pm$1.89 & 8.20$\pm$0.40 & 4.39$\pm$0.39 & 1.73$\pm$0.22 &  2.36$\pm$0.24 & 27.4$\pm$0.82 & 0.30$\pm$0.06 & 0.81$\pm$0.15 & 0.94$\pm$0.10  \\
NGC\,4748  & 7.39$\pm$0.37 & 73.57$\pm$2.84 & 7.85$\pm$0.61 & 1.90$\pm$0.42 & $<$0.32         &  1.59$\pm$0.20 & 11.32$\pm$0.67 &    0.40$\pm$0.13 & $<$0.35 & 0.94$\pm$0.09 \\
Mrk\,279   & 7.22$\pm$0.46 & 55.2$\pm$2.85 & 5.36$\pm$0.61 & 0.67$^{a}$    & 1.04$\pm$0.16 &  1.89$\pm$0.23 &  11.9$\pm$2.4 &   0.56$\pm$0.15 & 0.62$\pm$0.15 & 1.35$\pm$0.18\\
NGC\,5548  & 1.71$\pm$0.26 & 70.85$\pm$4.45 & 1.30$\pm$0.14 & $<$1.35  & $<$0.16  &  0.80$\pm$0.11 & 16.27$\pm$2.0 &   $<$0.14  & $<$0.11 & 1.31$\pm$0.24 \\
Mrk\,478   & $<$3.06 & 82.70$^{a}$ & ... &      3.73$\pm$0.40 & $<$0.90         &  1.26$\pm$0.15 & 6.90$\pm$0.48 &    ...      & $<$0.35  & ... \\
NGC\,5728  & 5.08$\pm$0.50 & 7.37$\pm$1.16 &  4.96$\pm$0.61 & 6.04$\pm$0.56 & 2.40$\pm$0.20 &  6.28$\pm$0.12 & 2.11$\pm$0.19 & 0.89$\pm$0.20 & 1.68$\pm$0.20 & 1.02$\pm$0.16  \\
PG\,1448+273& ... & ... & ... &  0.93$\pm$0.07 & 0.37$\pm$0.08 &  0.45$\pm$0.08 &  3.81$\pm$0.17 &      $<$0.20     & $<$0.20 & ... \\
Mrk\,291   & 0.91$\pm$0.24 & 4.27$\pm$0.44 & 0.80$\pm$0.09 & 0.30$\pm$0.12 & 0.22$\pm$0.07 &  0.47$\pm$0.10 & 1.68$\pm$0.19 &  $<$0.16 & $<$0.10 & 1.14$\pm$0.33 \\
Mrk\,493   & 1.96$\pm$0.34 & 3.03$\pm$0.59 & 1.40$\pm$0.27 &  ... & 0.32$\pm$0.07 &  1.04$\pm$0.20 & 3.45$\pm$0.25 &   $<$0.26 & $<$0.21 & 1.40$\pm$0.36 \\
PG\,1612+261&  3.56$\pm$0.25 & 50.92$\pm$0.98 & ... &    $<$1.02         & $<$0.07         &  0.76$\pm$0.17 &  ... &  ...       & ...  &  ...  \\
Mrk\,504   &  $<$0.72 & 21.8$\pm$1.22 & $<$0.20 &  $<$0.32 & 0.15$\pm$0.04 &  0.35$\pm$0.06 & 3.48$\pm$0.45  & $<$0.13     & 0.20$\pm$0.06  & ...  \\
1H\,1934-063& 7.24$\pm$0.24 & 159.0$\pm$1.51 & 7.46$\pm$1.05 &  2.38$\pm$0.65 & 1.80$^{a}$    &  1.23$\pm$0.20 & 18.6$\pm$0.72 & 0.31$\pm$0.10 & $<$0.10 & 0.97$\pm$0.14  \\
Mrk\,896   & $<$0.40 & 21.5$\pm$0.69 & $<$0.53 &  0.41$^{a}$    & 0.42$\pm$0.13 &  0.41$\pm$0.10 & 3.21$\pm$0.59 &   0.20$\pm$0.06 & $<$0.19 & ... \\
Ark\,564   & 3.87$\pm$0.44 & 90.9$\pm$1.50 & 3.90$\pm$0.65 &  0.80$\pm$0.22 & 0.53$\pm$0.18 &  1.24$\pm$0.27 & 9.74$\pm$0.52 &  $<$0.26  & $<$0.22 & 0.99$\pm$0.20 \\
\hline
\end{tabular}}

(a) Affected by telluric absorption. \\
(b) Total flux of the line.
\normalsize
\end{table*}

\begin{table*}
\begin{center}
\caption{Fluxes of the narrow component of the permitted \ion{H}{i} lines and
line ratios measured between \ion{H}{i}, \h2\ and \fe2.} \label{narrow}
\begin{tabular}{lcccc}
\hline \hline
Galaxy & Br$\gamma$  & Pa$\beta$ & \h2\ 2.12$\mu$m/Br$\gamma$ & \fe2\ 1.257$\mu$m/Pa$\beta$ \\
\hline
Mrk\,1210 &  6.01$\pm$0.37  & 31.45$\pm$0.88  & 0.33$\pm$0.03 & 0.50$\pm$0.03  \\
Mrk\,1239  & 8.40$\pm$1.14  & 77.16$\pm$1.65  & $<$0.07         & 0.10$\pm$0.01  \\
NGC\,3227  & 6.85$\pm$2.00  & 23.34$\pm$1.78  & 2.58$\pm$0.76 & 1.76$\pm$0.14  \\
NGC\,3310 &  10.79$\pm$0.33 & 31.74$\pm$1.03  & 0.14$\pm$0.03 & 0.33$\pm$0.03  \\
NGC\,4151 &  23.26$\pm$1.49 & 102.4$\pm$2.52  & 0.63$\pm$0.05 & 0.57$\pm$0.03  \\
Mrk\,766  &  7.40$\pm$0.53  &  27.15$\pm$0.47 & 0.32$\pm$0.04 & 0.28$\pm$0.03  \\
NGC\,4748 &  1.39$\pm$0.26  &  8.77$\pm$0.50  & 1.14$\pm$0.26 & 0.84$\pm$0.06  \\
NGC\,5548 &  1.07$\pm$0.16  &  3.76$\pm$0.45  & 0.74$\pm$0.15 & 0.45$\pm$0.09  \\
NGC\,5728 &  2.11$\pm$0.19  & 7.37$\pm$1.16   & 2.97$\pm$0.27 & 0.69$\pm$0.13  \\
PG\,1612+261 & 0.75$\pm$0.15& 3.23$\pm$0.18   & 1.01$\pm$0.30 & 1.10$\pm$0.10  \\
\hline
\end{tabular}
\end{center}
\end{table*}

In order to see how the fluxes listed in
Table~\ref{h2flux} compare with previous measurements
reported in the literature, we have listed in Table~\ref{liter}
fluxes for H$_{2}$ 2.121$\mu$m, [\ion{Fe}{ii}] 1.64$\mu$m,
Br$\gamma$, [\ion{Fe}{ii}] 1.257$\mu$m and Pa$\beta$ published
for a subsample of our objects along with the aperture size
used in the observations. If more than one measurement is reported 
for the same object, they are listed in different lines. This is 
to avoid mixing data taken under different observing conditions and 
setup. Clearly, the effect of the extended
emission is particularly important for NGC\,4151 and NGC\,5728.
For the remaining galaxies, most of the NIR  emission
seems to be concentrated in the circumnuclear region. It
important to keep in mind that the slit orientation may
vary for each entry of Table~\ref{liter}. However, due to the
relatively small apertures, its effects on the reported
fluxes may not be significant.

\begin{table*}
\begin{center}
\caption{Fluxes for the most important NIR lines taken from the literature. In units of
10$^{-15}$ erg cm$^{-2}$ s$^{-1}$} \label{liter}
\begin{tabular}{lcccccc}
\hline\hline

Galaxy   & H$_{2}$ 2.121$\mu$m & [\ion{Fe}{ii}] 1.64$\mu$m  & Br$\gamma$ & [\ion{Fe}{ii}] 1.257$\mu$m &  Pa$\beta$ \\
\hline
Mrk 1210  &  4.09$^{7}$        &  ...             & 7.11$^{7}$  & 77.1$^{7}$ & 37.4$^{7}$ \\
NGC\,3227 &     29$^{1}$       &   ...            &      ...     & ... & ...  \\
          &    ...             &  48$\pm$15$^{3}$ & 16.6$\pm$2.6$^{3}$ & ... & ...  \\
          &   14$\pm$0.7$^{6}$ &  32$\pm$1.7$^{6}$& 16$\pm$1.1$^{6}$  & ... & ...  \\
          & 11.6$\pm$0.08$^{8}$& 20.8$\pm$0.28$^{8}$ & 22.1$\pm$0.18$^{8}$ & ... & ... \\
NGC\,4051 &    8.7$^{1}$       &    ...           &     ...           & ... & ... \\
          & 4.2$\pm$0.5$^{6}$  & 2.0$\pm$0.5$^{6}$ & 13$\pm$1.9$^{6}$ & ...  & ... \\
NGC\,4151 &   2.0$^{1}$        &    ...           &    ...            & ...  & ... \\
          &    ...             &    ...           &  ...  &  36.7$\pm$9$^{4}$ & 836.2$^{4}$ \\
  & 50$\pm$5$^{5}$ & 180$\pm$10$^{5}$ & 160$\pm$10$^{5}$ & 230$\pm$10$^{5}$ & 2300$\pm$50$^{5}$ \\
Mrk\,766  &   2.4$^{1}$        &    ...           &    ...     &   ...    & ... \\
Mrk\,478  &    ...             &    ...           & 15.8$\pm$5$^{2}$  & ...  & ... \\
NGC 5728  &  2.6$\pm$0.4$^{6}$ &    ...           &    ...            & ...  & ... \\
          &  7.01$^{7}$        &    ...           & 2.22$^{7}$        & 29.9$^{7}$  & 21.8$^{7}$ \\
\hline
\end{tabular}
\end{center}
References - 1. Ruiz 1997 (1".2 slit width); 2. Rudy et al. 2001. (3" slit width).
3. Schinnerer et al. 2001. (3".6 circular aperture). 4. Knop et al. 1996
(0".75$\times$1".33 aperture). 5. Thompson 1995 (2"$\times$10" aperture). 6.
Sossa-Brito et al. 2001 (2" aperture). 7. Veilleux et al. 1997 (3"$\times$3" NGC\,5728,
6"$\times$3" Mrk\,1210 in J; 3"$\times$3" in K). 8. Reunanen et al. 2002 (1"x1".5).
\end{table*}

In the H band, we detected \h2\ (1-0)S(7) 1.747$\mu$m in NGC\,5728 and
NGC\,3227; it was the only \h2\ line detected in that spectral region.
This situation is in contrast with what is observed 
in planetary nebulae (Rudy et al.
\cite{rudy01}) and reflection nebulae (Martini et al. \cite{mar97}) for
example, where bright \h2\ lines from upper vibrational levels ($\nu
\geq$~2) are usually present.  In these sources UV fluorescence
dominates the production of H$_{2}$. In fact, the theoretical
predictions of Black \& van Dishoeck (\cite{bvd}) based on this
mechanism as the main source of energy input show that H$_{2}$ lines
such as (6-4)Q(1) 1.601$\mu$m and (5-3)Q(1) 1.493$\mu$m may have fluxes
comparable to that of (1,0)S(1). In our sample, the lack of
molecular hydrogen from rotational levels with $\nu >$~1 in the H band
suggests that UV fluorescence alone may have little importance
in the production of the observed H$_{2}$ spectrum. It is also possible
that these features are diluted by the strong continuum from the AGN
but this last hypothesis seems unlikely because [\ion{Fe}{ii}]
1.64$\mu$m is clearly visible in all objects and it often has an
intensity comparable to that of \h2\ (1,0)S(1).

Studies aimed at investigating the origin of \h2\ lines 
carried out on AGN samples show controversial
results. Veilleux et al.\ (\cite{vgh97}) claim that shocks associated
with nuclear outflows are a likely source of H$_{2}$, while Quillen et
al.\ (\cite{q99}) relate the H$_{2}$ emission to local star formation.
Our data can be used to put stronger constraints on this issue by
comparing the observed \h2\ fluxes with existing models that
discriminate between these two possible scenarios.  Mouri (\cite{mo94})
proposed that the line ratio \h2\ (1,0)S(2) 2.247$\mu$m/(1,0)S(1)
2.212$\mu$m separates shock-dominated gas (with I$_{2.247\mu
m}$/I$_{2.212\mu m}$=0.1-0.2) from fluorescent regions (with
I$_{2.247\mu m}$/I$_{2.212\mu m}$ $\sim$0.55). However, in dense gas
($n \geq$ 10$^{4}$ cm$^{-3}$) the collisional de-excitation of the \h2\
molecule must be taken into account; the spectrum is modified and
approaches the thermal spectrum seen in shocks.

\begin{figure*}
   \centering
   \includegraphics[width=15cm]{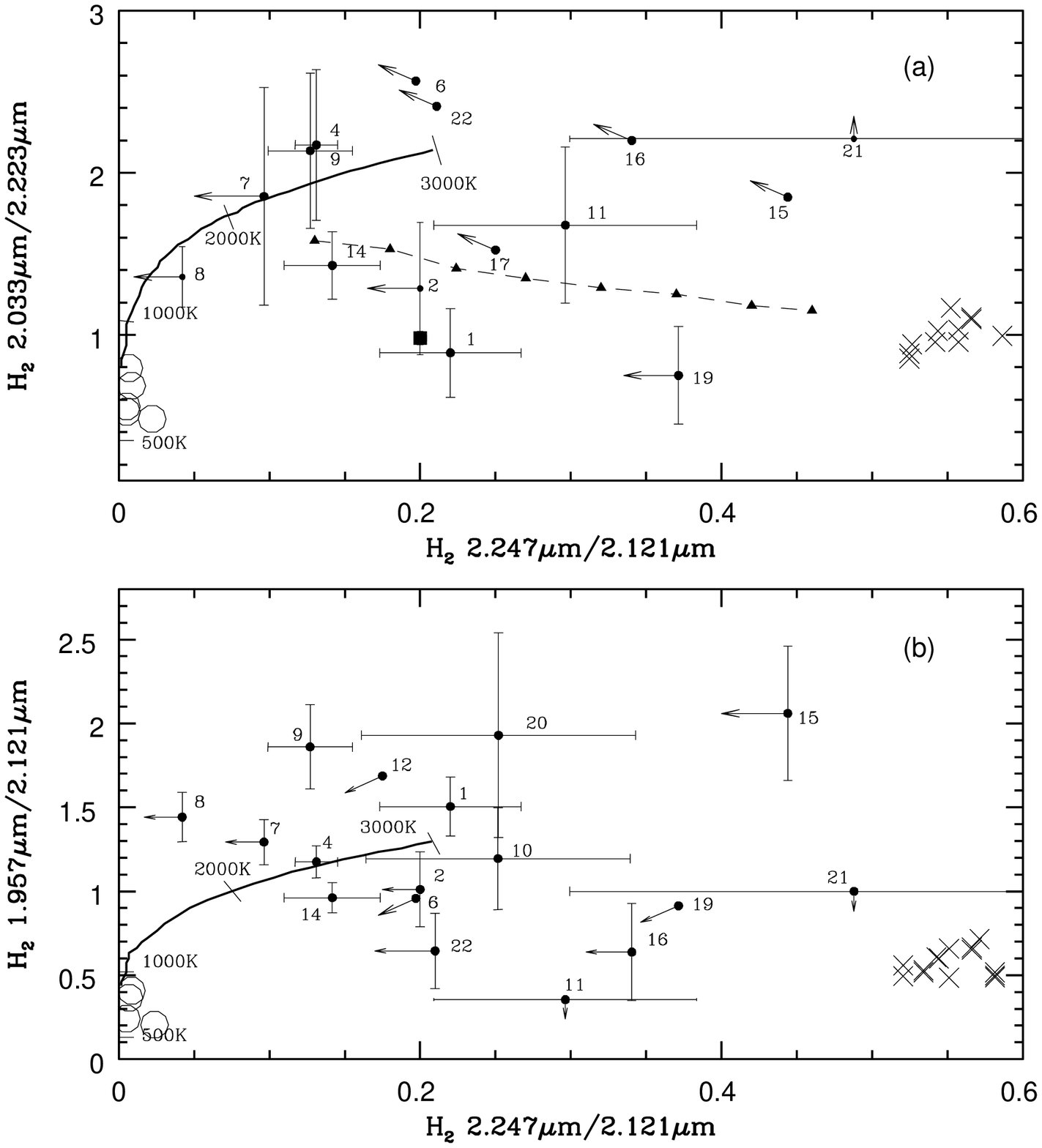}
      \caption{Plot of (a) (2-1)S(1) 2.247$\mu$m/(1-0)S(1) 2.121$\mu$m vs 
(1-0)S(2) 2.033$\mu$m/(1-0)S(0) 2.223$\mu$m and
(b) (2-1)S(1) 2.247$\mu$m/(1-0)S(1) 2.121$\mu$m vs 
(1-0)S(3) 1.957$\mu$m/(1-0)S(1) 2.121$\mu$m . Curves
represents thermal emission at 500-3000 K. Crosses are the non-thermal
UV excitation models of Black \& Van Dishoek (\cite{bvd}).
Open circles are thermal UV excitation
models of Sternberg \& Dalgarno (\cite{sd89}). The numbers identify each
object according to the notation given in column~1 of Table~\ref{log}.
The black square in (a) represents the datum for NGC\,253, a typical
starburst galaxy, taken from
Harrison et al. (\cite{ha98}). The full triangles, joined by a dashed-line,
represents the predicted line ratios from a mixture of thermal and
low-density fluorescence models of Black \& Van Dishoek (\cite{bvd}).
The percentage of the thermal component
decreases in steps of 10\% (and consequently the non-thermal component
increases by the same proportion), starting from a model where 90\%
is thermal and 10\% UV fluorescence (the first triangle from
left to right) up to a contribution of 20\% thermal and
80\% non-thermal (last triangle of the sequence, to the right).
              }
         \label{fig1}
   \end{figure*}

Figure~\ref{fig1}a shows (2,1)S(1) 2.247$\mu$m/(1,0)S(1) 2.212$\mu$m
plotted against (1,0)S(2) 2.033$\mu$m/(1,0)S(0) 2.223$\mu$m. The models
of Mouri (\cite{mo94}) are also shown.  These lines ratios are
reddening insensitive due to the very close wavelengths. A large
scatter is observed in  both line ratios for the galaxy sample.
NGC\,3227, NGC\,4051, NGC\,4151 and Mrk\,766 (labelled 4, 7, 8 and 9,
respectively) fall near the thermal curve model and indicate excitation
temperatures for the thermal component of between 1500\,K and 2500\,K.  For
these galaxies, the \h2\ emission can be considered purely thermal. For
the remaining objects, a mixture of thermal and non-thermal processes
are probably at work, with the excitation temperature of the thermal
component higher than 1000\,K. Interestingly, no AGN is located in the
region occupied by the pure non-thermal UV excitation models, except
possibly Mrk\,896.  However, its error bar makes any statement on this
object highly uncertain.  As expected, the starburst galaxy NGC\,3310
falls far from the locus of points for the non-thermal models. For
comparison, data for NGC\,253 (Harrison et al.\ \cite{ha98}), a
well-known starburst galaxy, are also plotted. NGC\,253 however, has a
thermal excitation temperature below 1000 K, similar only to Mrk\,1210
and Mrk\,504.  Although it is not possible to state at this far what
thermal mechanism excites the \h2\ lines, the results shown in
Figure~\ref{fig1}a tell us that the thermal component leads
over the non-thermal one.

In order to quantify the contribution of the thermal
excitation to the \h2\ lines, we
have also plotted in Figure~\ref{fig1}a the predicted
line ratios from a mixture of thermal (model S2) and
low-density fluorescence models (model 14) taken from
Black \& van Dishoek (\cite{bvd}).
These two models were empirically mixed in varying proportions
and the resultant emission line fluxes were then derived.
The calculated line ratios are represented by the dashed
line. It connects the models (represented
by the triangles) where the percentage of the thermal component
decreases in steps of 10\%, starting from a model where 90\%
is thermal and 10\% UV fluorescence (the first triangle from
left to right) up to a contribution of 20\% thermal and
80\% non-thermal. It can be seen that the observations
that are scattered across the plot may, in a first approach,
be explained by a suitable contribution of thermal and non-thermal
processes. Mrk\,504 is probably the AGN with the largest
contribution of the later mechanism.  In decreasing order
of importance, we find that UV fluorescence responds for
up to 50\% of the observed \h2\ spectrum in Mrk\,279,
30\% in Mrk\,1210, 25\% in Mrk\,124, and 15\% in
NGC\,5728.

Figure~\ref{fig1}b shows \h2\ (2-1)S(1) 2.247$\mu$m/(1,0)S(1)
2.212$\mu$m versus \h2\ (1-0)S(3) 1.963$\mu$m/(1-0)S(1) 2.212$\mu$m;
this is an additional diagnostic diagram proposed by Mouri
(\cite{mo94}), equivalent to that of Figure~\ref{fig1}a. However, it
has the drawback of including \h2\ (1-0)S(3) 1.957 $\mu$m, which may be
strongly affected by telluric absorption bands of H$_{2}$O and CO$_{2}$
if the redshift of the galaxy is between $z$=0.017--0.0398 or
it may be strongly blended with [\ion{Si}{vi}] 1.963$\mu$m. In addition,
if the broad component of Br$\delta$ is very conspicuous, it may
affect the determination of the continuum level around
\h2\ (1-0)S(3) 1.957 $\mu$m.  Overall,
Figure~\ref{fig1}b reproduces most of the conclusions already drawn:
NGC\,4051 and  NGC\,3227 fall on or near the pure
thermal excitation curve, and most objects show a thermal excitation
temperature in the range 1000--3000\,K indicating a dominance of thermal
processes over the non-thermal ones. The departure of Mrk\,766 and
NGC\,4151 from the thermal curve can be explained by the reasons stated
above.

\subsection{Rotational and vibrational temperatures}

The main result found in the previous section can be confirmed
by deriving rotational and vibrational
temperatures for the molecular \h2\ gas; thermal excitation must give
similar rotational and vibrational temperatures, as would be expected
for a gas in LTE.  Fluorescent excitation, on the other hand, is
characterized by a high vibrational temperature and a low rotational
temperature; non-local UV photons, not characteristic of the local
kinetic temperature, overpopulate the highest energy levels compared to
that expected for a Maxwell-Boltzmann population.  As a consequence, in
photodissociation regions where UV fluorescence dominates, the measured
vibrational temperature is $\sim$5000\,K whereas the rotational
temperature of the gas is $\sim$10 per cent of that value (Sternberg \&
Dalgarno \cite{sd89}).

The values of $T_\mathrm{vib}$ and $T_\mathrm{rot}$
found for the galaxy sample are listed in Table~\ref{temp}.
They were calculated using the fluxes of the observed
\h2\ lines listed in Table~\ref{h2flux} and the expressions
$T_\mathrm{vib} \cong 5600/\mathrm{ln}(1.355\times
I_\mathrm{1-0S(1)}/I_\mathrm{2-1S(1)})$ and  $T_\mathrm{rot} \cong
-1113/\mathrm{ln}(0.323\times I_\mathrm{1-0S(2)}/I_\mathrm{1-0S(0)})$
from Reunanen et al.\  (\cite{rka02}).  For Mrk\,1239, Mrk\,478,
H\,1143-192 and PG\,1612+261, neither of the two temperatures could be
derived either because only upper limits to the required fluxes are available
or because the lines were not observed.

Table~\ref{temp} strengths the hypothesis of
the \h2\ lines being purely thermal
in NGC\,3227, NGC\,4051, NGC\,4151
and Mrk\,766 because $T_\mathrm{vib} \simeq T_\mathrm{rot}$.
In Mrk\,1210, Mrk 504, NGC\,5728 and Mrk\,896, the two
temperatures differs appreciably,
indicating the presence of non-thermal excitation, as was
previously found from the sequence of mixed models of Figure~\ref{fig1}a.
The values listed in Table~\ref{temp} then confirms that for most
objects, a mixture of thermal and non-thermal processes contributes
to the observed \h2\ spectrum. Except for Mrk\,896
and Mrk\,504, it can be stated that thermal processes are
the leading excitation mechanism, with a contribution that
varies from 60\% up to 100\%.

The values for $T_\mathrm{vib}$ are also in very good agreement
to those derived by Reunanen et al.\ (\cite{rka02}) for a small sample
of AGN in which they found $T_\mathrm{vib}$ between 1800\,K and
2700\,K. For NGC\,3227, the only object common to both samples, a good
agreement is found for $T_\mathrm{vib}$ with our results
(2397$\pm$256\,K vs 1950\,K of Reunanen et al. \cite{rka02}).

\begin{table*}
\begin{center}
\caption{Vibrational and Rotational temperatures found for the
molecular hydrogen in the galaxy sample.} \label{temp}
\begin{tabular}{lccc|lcc}
\hline \hline
    &  $T_\mathrm{vib}$ & $T_\mathrm{rot}$ & & &  $T_\mathrm{vib}$ & $T_\mathrm{rot}$ \\
Galaxy & (K)          &      (K)       & & Galaxy & (K)          &      (K)  \\
\hline
Mrk\,1210   & 3100$\pm$700 & 900$\pm$300  & &  NGC\,5548    &  $<$2700 & ...  \\
Mrk\,124    & $<$2900        & 2500$\pm$800 & &  NGC\,5728  & 2500$\pm$600 & 1400$\pm$400\\
NGC\,3227   & 2400$\pm$300 & 3100$\pm$700 & &  PG\,1448+273 &  $<$5000 & $<$2200 \\
NGC\,3310   & $<$3000      & $<$2200      & &  Mrk\,291     &  $<$4100 & $<$3300 \\
NGC\,4051   & $<$2100      & 2200$\pm$800 & &  Mrk\,493     &  $<$3300 & $<$1600 \\
NGC\,4151   & $<$1600      & 1300$\pm$200 & &  Mrk\,504     &  $<$4300 & 800$\pm$300 \\
Mrk\,766    & 2400$\pm$500 & 3000$\pm$700 & &  1H\,1934-063 &  3300$\pm$1200 & ... \\
NGC\,4748   & 3300$\pm$1200&  ...         & &  Mrk\,896     &  5500$\pm$2100 & $<$3300 \\
Mrk\,279    & 3700$\pm$1100& 1800$\pm$500 & &  Ark\,564     &  $<$3000 & $<$4400 \\
\hline
\end{tabular}
\end{center}
\end{table*}

If the main excitation mechanism for \h2\ in AGN is thermal,
it is interesting to examine the possible emission mechanisms as
either shocks (produced by supernovae winds or by a radio
jet), or UV or X-ray heating are potential candidates for the \h2\ emission.

Recently, Rodr\'{\i}guez-Ardila \& Viegas (\cite{rv03}) found evidence
of starburst activity in Mrk\,766, NGC\,4051, and NGC\,3227 by
observing the 3.3$\mu$m polycyclic aromatic hydrocarbon (PAH) emission
in these objects.  The beam size covered by their spectra is similar to
the one used here.  Galaxies with high levels of star formation
activity feature strong PAH emission and prominent \h2\ lines.
The relationship between these two characteristics was initially
established by Mouri et al. (\cite{mo90}) and later confirmed by
Mizutani et al. (\cite{mi94}), who explained it
in terms of the physical processes in photodissociation regions
associated with relatively massive young stars. Our data
provides further support to the hypothesis of the starburst as the
origin of the observed molecular hydrogen emission in those three objects.  In
this scenario, shocks driven by the supernova remnants appears as the
most likely source for the \h2\ excitation. In an AGN with a starburst
component, part or all of the observed molecular emission is expected
to arise from this process.  This seems to be the case of Mrk\,766,
NGC\,4051, and NGC\,3227. The predicted \h2\ 2-1$S$(0)/1-0$S$(1) ratio
for the pure shock model (S2) of Black \& van Dishoek
(\cite{bvd}) at $T$=2000~K is
0.082, in marginal agreement to the measured ratios of 0.13$\pm$0.03,
0.096, and 0.13$\pm$0.01, respectively. Their S2 model also predicts a
\h2\ 1-0$S$(0)/1-0$S$(1) ratio of 0.21, well in accord to the values of
0.34$\pm$0.07, 0.26$\pm$0.07, and 0.21$\pm$0.04, respectively.

Though supernova-driven shocks offer a plausible scenario to explain
the pure thermal \h2\ spectrum observed in those three objects, we
cannot ignore that AGN are powerful sources of X-ray emission. It
is then expected that X-ray heating may have a non-negligible contribution
to the observed \h2\ flux. Moreover, supernova remnants themselves
emits high-energy radiation capable of heating the gas.

For NGC\,4151, the non-detection of the 3.3 $\mu$m feature
within the inner 3''.8$\times$3.''8 region (Imanishi et al.
\cite{ima98}) makes unlikely the hypothesis of a starburst
as the main contributor of the \h2\ lines. Further efforts
to detect signatures of starburst activity in that Seyfert 
have failed (e.g.\ Sturm et al. \cite{stu99}). The analysis
of the molecular line profiles carried out in \S~\ref{kin}
showed that it is unlikely that \h2\
is produced by shocks from the radio jet due to the small value
of FWHM measured. Thus X-ray
heating is the most plausible mechanism for
\h2\ excitation in this object; specific models to test this
possibility are discussed in \S~\ref{x-ray}.

Finally, we comment on Mrk\,1210. For this galaxy, we derived a
$T_\mathrm{rot} \sim$ 900 K. We already know that UV fluorescence
has at most, a 30\% contribution.
It means that the low $T_\mathrm{rot}$ can be attributed to UV
heating and this process can be the main source of \h2\ excitation.
UV heating models (Black \& van Dishoek \cite{bvd}),
where FUV photons emitted by OB stars (photodissociation regions) heat
the gas, predict excitation temperatures of $\leq$1000K for the
\h2\ thermal component. Such low values are typically found in
starburst galaxies (Mouri \cite{mo94}), suggesting that the
\h2\ emission observed in Mrk\,1210 may  primarily come
from a circumnuclear starburst. In fact, the position of Mrk\,1210
in Figure~\ref{fig1}a is almost coincident with that of NGC\,253,
a well-known starburst galaxy. Additional evidence giving
support to this hypothesis was found in Section~\ref{kin},
where the kinematics of the \h2\ gas pointed out that it was
unrelated to the AGN. Further arguments
supporting this view come from Storchi-Bergmann et al. (\cite{sb98})
and Schulz \& Henkel (\cite{sh03}) who reported the detection of a
Wolf-Rayet (WR) feature around $\lambda$4686\,\AA\ in the inner 2''
of Mrk\,1210. This aperture coincides with the beam size of our
observation. Moreover, WR-features are taken as unambiguous signs for the
presence of a starburst
(Schulz \& Henkel \cite{sh03}) and due to the close
relationship between \h2\ and starburst activity (Mouri \cite{mo90};
Mizutani et al. \cite{mi94}) it is natural to attribute the
\h2\ emission to this last component. Observations at higher
spatial resolution can test this scenario.

In summary, the observational evidence presented up to now
shows that the molecular hydrogen lines in AGN are
mostly thermal, with a contribution of non-thermal processes
typically  varying from 15\% to 30\%.  There is not unique
process associated with the thermal component. \h2\ gas
heated by UV photons emitted from photodissociation regions,
X-rays from the AGN, and shocks from supernovae remnants
arise as the most likely sources of \h2\ excitation. Our
results agree with those of Rigopoulou et al. (\cite{ri02}),
found using mid-infrared \h2\ lines from ISO observations
for a sample of Seyfert~2 galaxies.

\section{The origin of the \fe2\ Lines} \label{xray}

The origin of [Fe\,{\sc ii}] in active galaxies is highly
controversial. It can be produced by shock excitation from the radio
jets and/or photoionization from the central source. In addition, fast
shocks associated with supernova remnants can be an additional source
of [Fe\,{\sc ii}], as is fairly well established in starburst galaxies
(Simpson et al. \cite{sfbw96}; Moorwood \& Oliva \cite{mo88}).
As found by Forbes
\& Ward (\cite{fw93}), the strength of the
[Fe\,{\sc ii}] emission in the central regions of AGN is tightly
correlated with the 6-cm radio emission, and that the same correlation
is obeyed for both starburst and Seyfert galaxies.  This was taken as
evidence that the most likely mechanism for the production of [Fe\,{\sc
ii}] emission in Seyfert galaxies is shock excitation from the radio
jets and/or SNRs.

From Table~\ref{h2flux} and Figures~\ref{fig1a} to~\ref{fig1c}, it can
be seen that \fe2\ 1.257$\mu$m and \fe2\ 1.64 $\mu$m were detected in 
all objects of our sample but PG\,1448+273. In the latter 
AGN, both lines are located in regions of very poor atmospheric
transmission, hampering any effort to detect them. 

The flux ratios \fe2\ 1.644$\mu$m/Br$\gamma$, or equivalently,
\fe2\ 1.257$\mu$m/Pa$\beta$\footnote{The flux ratios
\fe2\ 1.644$\mu$m/Br$\gamma$ and \fe2\ 1.257$\mu$m/Pa$\beta$ are
equivalent because the two \fe2\ lines share the same upper level. They
are related by the equation \fe2~1.257$\mu$m/Pa$\beta$ =
0.231*\fe2~1.644$\mu$m/Br$\gamma$.  The advantage of using
\fe2\ 1.257$\mu$m/Pa$\beta$ instead of \fe2\ 1.644$\mu$m/Br$\gamma$ is
that the former is reddening insensitive due to the proximity in
wavelength of both lines, serving as a more reliable diagnostic.} can
be used to study the origin of the \fe2\ emission because there is an
increasing progression in the ratio from pure photoionization (as in
\ion{H}{ii} regions) to pure shock excitation (i.e.\ supernovae
remnants), with starburst and active galaxies located at intermediate
values (Mouri et al. \cite{mkt93}; Goodrich et al.  \cite{gvh94};
Alonso-Herrero et al.  \cite{alh97}). In starburst galaxies the ratio
[Fe\,{\sc ii}] 1.644$\mu$m/Br$\gamma$ is observed to be in the range
0.5$-$2 (Simpson et al. \cite{sfbw96}), in reasonable agreement with
the starburst models of Colina (\cite{co93}) which predicts ratios
between 0.1 and 1.4. In these objects, the \fe2\ emission originates in
supernova-driven shocks (Mouri et al.  \cite{mkt00}).  AGN usually show
larger values of \fe2/Br$\gamma$ than starburst galaxies, probably
reflecting the effect of the Seyfert nuclei (Kawara et al.
\cite{knt98}). In fact, X-ray emission, which is dominant in Seyferts,
can penetrate deeply into atomic gas and create extended partly ionized
regions where \fe2\ can be formed. Models presented by Alonso-Herrero
et al. (\cite{alh97}) show that X-rays are able to explain
\fe2/Br$\gamma$ ratios up to $\sim$20.

Col.~5 of Table~\ref{narrow} lists the flux ratio [Fe\,{\sc ii}]
1.257$\mu$m/Pa$\beta$ as derived for our galaxy sample. Only the
objects in which it was possible to carry out a clean deblending of the
narrow component of the permitted lines are included in this analysis.
Overall, our line ratios agree with the values found for other samples
of Seyfert galaxies, 0.5 $<$ \fe2 1.257$\mu$m/Pa$\beta <$ 4.6 (see
Fig.\,3 of Alonso-Herrero et al. \cite{alh97}).  Interestingly,
Mrk\,766, a NLS1 with a circumnuclear starburst, has a \fe2
1.257$\mu$m/Pa$\beta$ ratio similar to that observed in starburst
galaxies, indicating that most of \fe2\ we detect is of stellar
origin.  On the other hand, NGC\,3227, a classical Seyfert~1 also
harboring a circumnuclear starburst, has a high
\fe2\ 1.257$\mu$m/Pa$\beta$ ratio, pointing towards an AGN origin.
NGC\,4151 and NGC\,5548 are in the borderline of the starburst-AGN
division but any contribution from star forming regions would be
negligible because of the lack of circumnuclear starburst in these
objects.

\begin{table*}
\begin{center}
\caption{Line ratios \h2\ 2.12$mu$m/Br$\gamma$ and \fe2\ 1.257$\mu$m/Pa$\beta$,
from AGN taken from the literature.} \label{fe2lit}
\begin{tabular}{lccc}
\hline \hline
 Galaxy   &      \h2\ 2.121$\mu$m/Br$\gamma$ & \fe2\ 1.257$\mu$m/Pa$\beta$ & Reference \\
\hline
NGC\,1386 &   1.14$\pm$0.10 & 1.59$\pm$0.14  & 1 \\
NGC\,4945 &   3.13$\pm$0.06 & 0.86$\pm$0.03  & 1 \\
NGC\,5128 &   2.04$\pm$0.10 & 3.42$\pm$0.16  & 1 \\
NGC\,1097 &   $>$4.40       & $>$0.43        & 1 \\
Mrk\,1066 &   0.89$\pm$0.06 & 0.75$\pm$0.04  & 2 \\
NGC\,2110 &   3.17$\pm$0.63 & 8.11$\pm$0.78  & 2 \\
NGC\,4388 &   0.88$\pm$0.08 & 0.40$\pm$0.03  & 2 \\
Mrk\,3    &   0.31$\pm$0.04 & 1.24$\pm$0.04  & 2 \\
\hline
\end{tabular}
\end{center}
References: (1) Reunanen et al. (\cite{rka02}); (2) Knop et. al. (\cite{knop01})
\end{table*}

Although the \fe2\ 1.257$\mu$m/Pa$\beta$ is helpful to discriminate
between a stellar and non-stellar origin for \fe2, it can say little
about the excitation mechanisms that can lead to the \fe2\ lines in
the latter case.  Heating either by X-rays or shocks created by mass
outflows from the nuclei (i.e., jets and/or winds interactions with
ambient clouds) are plausible scenarios.
As proposed by Mouri et al.\ (\cite{mkt00}), these two
processes can be discriminated by the electron temperature of the
\fe2\ region: 8000\,K in heating by X-rays, and 6000\,K in heating by shocks.
Determining the \fe2\ temperature, however, is a difficult
task, mostly because some of the diagnostic lines are intrinsically
weak and heavily blended with broad adjacent features. The alternative
is to examine if X-rays can explain the observed emission. This issue
is addressed in detail in Section~\ref{x-ray}.

\subsection{Are \fe2\ and \h2\ related?}

The observational evidence presented throughout this work indicates
that \fe2\ and \h2\ are common features in the nuclear spectra of AGN,
although probably originated from different parcels of
gas. However, this does not exclude the possibility that both set of
lines originate from a single, dominant mechanism.  Moreover, if
dust is intermixed with the line emitting clouds, and there are strong
velocity gradients, \fe2\ and \h2\ lines can appear to have different
velocity fields (Knop et al. \cite{knop01}) even though they
are produced in adjacent regions.

\begin{figure}
   \centering
   \includegraphics[width=10cm]{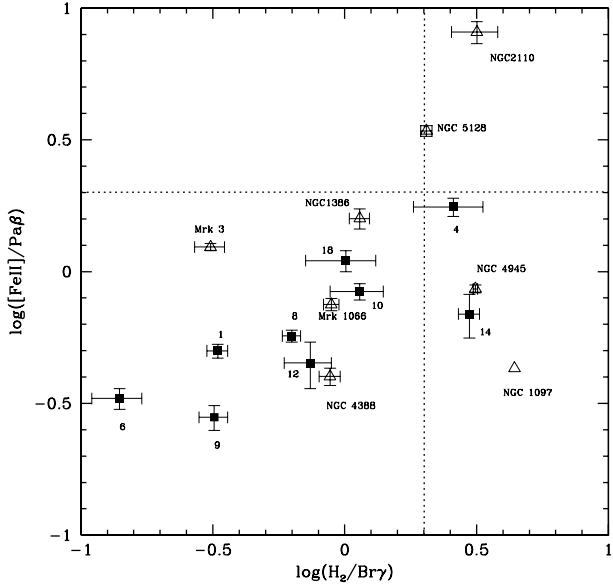}
      \caption{The ratios \h2\ 2.121$\mu$m/Br$\gamma$ and
\fe2\ 1.257$\mu$m/Pa$\beta$. Only the Seyfert~1s and NLS1 in which
a reliable deblending of the narrow component of the \ion{H}{i} lines
was carried out were plotted. Filled circles represents our
data and open symbols data taken from the literature. Triangles are
measurements from Knop et al. (\cite{knop01}) and squares from
Reunanen et al. (\cite{rka02}); see text for further details. The
number along each data point identifies our galaxies according to
the notation given in column~1 of Table~\ref{log}.}
         \label{he2fe2c}
   \end{figure}

In order to examine a possible relationship between  \fe2\ and \h2,
we have plotted in Fig.~\ref{he2fe2c} \h2\ 2.121$\mu$m/Br$\gamma$ versus
\fe2\ 1.257$\mu$m/Pa$\beta$ for our sample along with
data taken from the literature for Seyfert~2 galaxies observed
with a similar beam size and resolution (see Table~\ref{fe2lit}).
Note that the line ratios are completely reddening free due
to their close proximity in wavelength for the lines involved.

Interesting results are seen in Fig.~\ref{he2fe2c}. For most
objects with both \h2/Br$\gamma$ and \fe2/Pa$\beta<$2, the two
ratios shows a possible, positive correlation. However, for values of either
ratio larger than 2, the correlation breaks down and no trend
is observed. In addition, no Seyfert~1 galaxy displays extreme
values of any of the line ratios plotted in Figure~\ref{he2fe2c};
Both \h2/Br$\gamma$ and \fe2/Pa$\beta >$2 occurs predominantly
in Seyfert~2 galaxies.

The correlation observed in Fig.~\ref{he2fe2c} is
consistent with the one found by Larkin et al. (\cite{lar98}) in a study of
LINERs and other emission line objects, including a few Seyferts. They report a
strong linear correlation in the log-log space between \fe2/Pa$\beta$ and \h2/Br$\gamma$,
with LINERs displaying the largest ratios and starbust galaxies the smallest
ones.  Larkin et al. (\cite{lar98})
suggested that \fe2/Pa$\beta \sim$1 and \h2/Br$\gamma \sim$3 marked
the end of Seyfert-like nuclei and the beginning of LINER-like
objects. The larger sample of Seyfert galaxies employed here
allow us to propose that the borderline between Seyfert and LINER
activity occurs at \fe2/Pa$\beta \sim$2 and \h2/Br$\gamma \sim$2.
In this sense, Fig.~\ref{he2fe2c} would serve as a diagnostic
diagram in the NIR to separate galaxies according to their
level of nuclear activity: starburts occupying the bottom left corner of
the plot, with \fe2/Pa$\beta$ and \h2/Br$\gamma <$0.3, Seyferts
occupying the locus of points with values of either ratio
between 0.4--2, and LINERS showing ratios larger than 2.
We note that, as in our plot, Larkin et al. (\cite{lar98})
found that some of the Seyfert~2 galaxies lie significantly below
or above the correlation.

Assuming that \fe2/Pa$\beta$ and \h2/Br$\gamma$ are in fact correlated,
interpreting the origin of that correlation is non-trivial. This is 
mainly because Fig.~\ref{he2fe2c} does not
discriminate between the different processes that may give rise to
either \fe2\ or \h2\ in each object. The most straightforward
interpretation is that both set of lines are excited by the same
mechanism. It is yet
possible that more than one mechanism drives the production of 
\fe2\ and \h2\ within the same object, but environmental
effects dilutes the effects of one mechanism over the other
(i.e., the presence of a circumnuclear starburst in an AGN).

Recalling that the objects plotted in  Fig.~\ref{he2fe2c} are all 
AGNs, it is natural to consider that the dominant excitation mechanism 
discussed above is related to the central engine. In particular,
X-rays from the nucleus can heat the NLR gas through the large 
partially ionized zone it creates, driving the \fe2\ and \h2\
emission. Since 98\% of the iron is tied up in dust grains, this 
process must free up iron through dust destruction
and yet not destroy the \h2\ molecules. Blietz et al. (\cite{bli94})
and Knop et al (\cite{knop01}) show that hard X-rays can do it for a
sample of Seyfert 2 galaxies.
Additional evidence of the role of X-rays can be drawn from 
the good correlation between \h2\ 2.121$\mu$m/Br$\gamma$ and 
[O\,{\sc i}] 6300\AA/H$\alpha$ and between \fe2\ 1.257$\mu$m/Pa$\beta$ 
and  [O\,{\sc i}] 6300\AA/H$\alpha$ reported by Larkin et al. (\cite{lar98}).
They indeed argue that hard X-ray heating from a power-law source is 
a plausible mechanism to explain LINERS with low values of the 
\fe2/Pa$\beta$ and  \h2\ 2.121$\mu$m/Br$\gamma$ ratio, to which they 
suggest, are low-luminosity Seyfert galaxies. This same picture can 
be extended to the Seyfert galaxies of our sample to support the 
slim correlation reported here.

\subsection{X-rays as a source of \fe2\ and \h2\ lines} \label{x-ray}

In order to test if the observed \fe2\ and \h2\ emission can be
attributed to X-ray heating, we have calculated the emergent
\h2\ 2.121$\mu$m and \fe2\ 1.644$\mu$m fluxes using the models of X-ray
excitation of Maloney et al. (\cite{mht96}). These models predict the
resultant infrared emission (lines and continuum) of a neutral gas
cloud of electron density $n_\mathrm{e}$ exposed to a source of hard X-rays
of brightness $L_\mathrm{x}$ located at a distance $d$ from the
cloud. The main parameter controlling the physical state of the
X-ray-illuminated gas is the effective ionization parameter,
$\zeta_\mathrm{eff}$, which is the attenuated X-ray flux to density
ratio, determined from the expression
\begin{equation}
\zeta_\mathrm{eff}~\simeq~100\frac{L_\mathrm{44}}{n_\mathrm{5}d_\mathrm{pc}^\mathrm{2}N_\mathrm{22}^\mathrm{0.9}}
\,,
\end{equation}
where $L_\mathrm{44}$ is the hard X-ray luminosity,
$L_\mathrm{x}$, in units of 10$^{44}$ erg~sec$^{-1}$, $n_\mathrm{5}$ is
the gas (total hydrogen) density in units of 10$^{5}$ cm$^{-3}$,
$d_\mathrm{pc}$ is the distance, in parsecs, to the AGN from the
emitting gas, and $N_\mathrm{22}$ is the attenuating column density
between the AGN and the emitting gas, in units of 10$^\mathrm{22}$
cm$^{-2}$. Once $\zeta_\mathrm{eff}$ is computed, the predicted
\h2\ and \fe2\ emergent intensities can be obtained with the help of
Figs.~6a and~6b of Maloney et al. (\cite{mht96}) for a gas density of
10$^{5}$ cm$^{-3}$ and 10$^{3}$ cm$^{-3}$, respectively.

The input parameters employed in our modeling are listed in
Cols.~2--4 of Table~\ref{x-rayh2}. Two values of the hydrogen
column density, N$_{\mathrm{H}}$, are listed and were used in the modeling:
one determined directly from X-rays (Col.~3) and one from 21~cm radio
observation (Col.~4). If this last value was not available,
we used instead the N$_{\mathrm{H}}$ derived from E(B-V)
according to the expression H$_\mathrm{\ion{H}{i}}$ = 5.2
$\times$ 10$^{21}$ E(B-V) cm$^{-2}$ (Shull \& van Steenberg
\cite{svs85}), which assumes Galactic dust/gas ratio. The
distance {\it d} between the X-ray source and \ion{H}{i}
cloud was assumed to be equal to the radius of the integrated
region (see Col~10 of Table~\ref{log}). This value represents
the maximum distance at which the line could be emitted.

As a first test, we determined the predicted \h2\ 2.121$\mu$m
intensity for a gas density $n_{\mathrm{e}}$=10$^{5}$~cm$^{-3}$.
The results appears in Cols.~6 and~7 of Table~\ref{x-rayh2},
for N$_{\mathrm{H}}$ derived from X-rays or either radio observations or
E(B-V), respectively. Compared to the observations, \h2\
intensities calculated using N$_{\mathrm{H}}$ obtained from
X-ray data offer a poorer
fit. This can be understood if we take into account
that the hydrogen column density derived from hard X-rays maps
material obscuring the innermost part of active galaxies, while the
\h2\ and \fe2\ emission is expected to arise farther out in the NLR.
For this reason, we opted to use the N$_{\mathrm{H}}$ values
determined either from 21\,cm measurements or E(B-V).  With these constraints,
we calculated the emergent \fe2\ and \h2\ intensities and the
results are listed in Table~\ref{x-rayfe2}.

\begin{table*}
\begin{center}
\caption{Predicted \h2\ 2.12$\mu$m fluxes from the X-ray models of Maloney et al. (\cite{mht96}) for a density of 10$^{5}$ cm$^{-3}$.} \label{x-rayh2}
\begin{tabular}{lccccccc}
\hline \hline
  & log $L_\mathrm{x}^{1}$ & $N_\mathrm{Hx}$ & $N_\mathrm{H,NLR}$ &    & \multicolumn{2}{c}{Predicted \h2\ 2.12$\mu$m} \\
\cline{6-7}
Galaxy & (erg~s$^{-1}$) & (10$^{21}$~cm$^{-2}$) & (10$^{21}$~cm$^{-2}$) & $F_{2.12\mu m}^{\mathrm{2}}$ & $N_\mathrm{Hx}^{2}$ & $N_\mathrm{H~NLR}^{2}$ \\
(1) & (2) & (3) & (4) & (5) & (6) & (7) \\
\hline
Mrk\,1210 & 42.36 & 0.83 & 0.286$^{3}$ & 2.00$\pm$0.12 & 0.32 & 0.58 \\
NGC\,3227 & 42.00 & 0.50 & 0.036 & 17.7$\pm$1.00 & 1.29 & 0.68 \\
NGC\,4051 & 41.20 & 6.30 & 0.156$^{3}$ & 5.81$\pm$0.37 & 0.16 & 1.51 \\
NGC\,4151 & 43.00 & 2.26 & 0.380 & 14.7$\pm$0.55 & 1.45 & 24.6 \\
Mrk\,766  & 43.00 & 1.28 & 0.260$^{3}$ & 2.36$\pm$0.24 & 0.67 & 1.68 \\
NGC\,5548 & 43.76 & 4.57 & 0.021 & 0.80$\pm$0.11 & 0.74 & 283.0 \\
NGC\,5728 & 41.34 & ...  & 0.391$^{3}$ & 6.28$\pm$0.12 & ...  & 0.19 \\
\hline
\end{tabular}
\end{center}
Notes. Col(1) $-$ Object name, Col.(2) $-$ Logarithmic 2-10 keV luminosity,
Col.(3) $-$ hydrogen column density determined from X-ray observations,
Col.(4) $-$ hydrogen column density for the NLR determined either
from radio 21~cm observations or the E(B-V).
Col.(5) $-$ Observed \h2\ 2.12$\mu$m flux, Cols.(6-7) $-$ predicted
\h2\ 2.12$\mu$m flux using column density from X-ray and radio, respectively.\\
$^{\mathrm{1}}$ {\it ASCA} 2--10 keV luminosities were taken from George et al.
(\cite{geo98}) for NGC\,3227, NGC\,4051, NGC\,4151, Mrk\,776 and NGC\,5548 and
from Awaki et al. (\cite{aw00}) for Mrk\,1210.\\
$^{\mathrm{2}}$ In units of 10$^{-15}$~erg~s$^{-1}$~cm$^{-2}$ .\\
$^{\mathrm{3}}$ Determined from the relation N$_\mathrm{\ion{H}{i}}$=5.2$\times$10$^{21}$E(B-V). See text for further details.
\end{table*}

\begin{table*}
\begin{center}
\caption{X-ray model parameters and predicted line fluxes for \fe2\ 1.64$\mathrm{\mu}$m and \h2\ 2.12$\mathrm{\mu}$m.} \label{x-rayfe2}
\begin{tabular}{lccccccc}
\hline \hline
  & $n_\mathrm{e}$ & $N_\mathrm{H}$& \multicolumn{2}{c}{\fe2\ 1.64~$\mathrm{\mu m}^{1}$} & & \multicolumn{2}{c}{\h2\ 2.12~$\mathrm{\mu m}^{1}$} \\
\cline{4-5} \cline{7-8}
Galaxy & (cm$^{-3}$) & (10$^{21}$~cm$^{-2}$) & Obs. & Predicted & & Obs. & Predicted \\
\hline
Mrk\,1210 & 10$^{5}$ & 0.286 & 17.1$\pm$0.64 & 0.0  & & 2.00$\pm$0.12 & 0.58 \\
NGC\,3227 & 10$^{3}$ & 0.036 & 41.0$\pm$3.60 & 7.08 & & 17.7$\pm$1.00 & 33.9 \\
NGC\,4051 & 10$^{5}$ & 0.156 & 6.42$\pm$0.97 & 0.0  & & 5.81$\pm$0.37 & 1.51 \\
NGC\,4151 & 10$^{3}$ & 0.380 & 56.2$\pm$1.96 & 10.7 & & 14.7$\pm$0.55 & 24.6 \\
Mrk\,766  & 10$^{5}$ & 0.260 & 8.20$\pm$0.40 & 0.0  & & 2.36$\pm$0.24 & 1.68 \\
NGC\,5548 & 10$^{3}$ & 0.021 & 1.30$\pm$0.14 & 15.8 & & 0.80$\pm$0.11 & 10.96 \\
NGC\,5728 & 10$^{5}$ & 0.391 & 4.96$\pm$0.91 & 0.0  & & 6.28$\pm$0.12 & 0.19 \\
\hline
\end{tabular}
\end{center}
$^{\mathrm{1}}$ In units of 10$^{-15}$~erg~s$^{-1}$~cm$^{-2}$.
\end{table*}

The results presented in
Table~\ref{x-rayfe2} suggest that X-ray heating is responsible
for only a fraction of the \fe2\ and \h2\ emission, except in
NGC\,5548 where it can fully drive it. In four out of seven objects,
the predicted \fe2\ flux is negligible while in NGC\,4151 and
NGC\,3227, X-rays may contribute for up to 15\% of the measured \fe2. Regarding \h2,
X-rays contribute to the observed emission but only in NGC\,4151
NGC\,5548, and NGC\,3227 can it entirely explain the excitation of the
molecular gas. Note that for the Seyfert~2 galaxies Mrk\,1210 and
NGC\,5728, an additional source of \h2\ would be necessary because
X-rays contribute with only a small fraction of the measured flux.

The above discrepancy can be alleviated if the emitting 
clouds were effectively located closer to the X-ray source than the 
conservative distance we
adopted in the models (equals to the beam size of
our observations). In Mrk\,1210, for instance, neutral clouds with
$n_{\mathrm{e}}$=10$^{3}$~cm$^{-3}$ at $\sim$80~pc can generate
enough \h2\ and \fe2\ to account for the observations. A
similar result is obtained for NGC~5728. This possibility would
obviate the need of additional excitation mechanisms and readily
explain the correlation observed in Fig.~\ref{he2fe2c}. 
In order to confirm or discard this scenario, data taken at a superior
resolution in order to map the inner molecular gas and constrain
its density and distance from the central source are necessary. 
For the time being, we are only in a position to
state that overall, the role of X-ray heating in exciting \h2\ and \fe2\
lines in AGN is confirmed but  the strength at which it operates
strongly varies from object
to object.

Morphological effects, more visible in Seyfert~2 than in
Seyfert~1 galaxies, can be behind the strong departure from the
correlation of Fig.~\ref{he2fe2c} presented by some objects of
the sample. In effect, most of the galaxies with large
\h2\ 2.121$\mu$m/Br$\gamma$ or \fe2\ 1.257$\mu$m/Pa$\beta$ ratios
share similar morphological aspects. For example, the presence of
ionization cones and, most importantly, dust lanes, favour the
production of \h2\ over \fe2.  This could be the case in NGC\,4945 and
NGC\,5728, where such features are clearly detected (Reunanen et al.
\cite{rka02}; Wilson et al.  \cite{wil93}).  In these objects,
\h2\ 2.121$\mu$m/Br$\gamma>$2. Jets and outflows, on the other hand,
tend to favour \fe2\ emission over \h2. This is the case in the
galaxies NGC\,2110 (which shows one of the \fe2/Pa$\beta$ ratios ever
measured in an AGN, Knop et al. \cite{knop01}), NGC\,5128
(Centaurus\,A), and Mrk\,3. In this case,
\fe2\ 1.257$\mu$m/Pa$\beta>$2. Whether these features are in fact
related to the enhancement of any of the two emissions
cannot be deduced from our data. Higher spatial resolution is needed to
address this question.

Finally, it is interesting to note that X-rays models can explain
\fe2/Br$\gamma \sim$20, as found for Alonso-Herrero et al. (\cite{alh97}),
but it would need quite a high X-ray flux or clouds located
too close to the central source. None of the objects studied (except probably
NGC~5548), has the right combination of X-ray emission, hydrogen
column density, and distance from the X-ray source to the emitting clouds,
to produce ratios as high as 20 (using Maloney's models). Since these
variables were constrained from observations, the results reflect
the conditions for the galaxies of our sample.

\subsection{The \fe2\ lines as useful indicators of reddening for the NLR}

The simultaneous measurements of \fe2\ 1.257$\mu$m and
1.644$\mu$m allow us to test, for the first time on such a
big sample of AGN, how theory compares to observation.
The ratio \fe2\ 1.257$\mu$m/1.644$\mu$m
is fixed by atomic physics to be 1.34 (Bautista \& Pradhan \cite{bp98})
since both lines originate from the same upper level (i.e.
$^{4}D_{7/2}$-$^{6}D_{9/2}$/$^{4}D_{7/2}$-$^{4}F_{9/2}$).
It means that this ratio is not only a good check on theoretical
predictions but also a reliable indicator of the extinction towards the
NLR because it is independent of temperature and density.
Last column of Table~\ref{h2flux} lists the
\fe2~1.257$\mu$m/1.644$\mu$m ratio measured in each galaxy.
All but Mrk~279, NGC~5548 and Mrk~493,
tend to have ratios $\sim$30\% lower than the theoretical
prediction. In fact, the mean \fe2\ 1.257$\mu$m/1.644$\mu$m of
the sample equals 0.98. The uncertainties in the observed line
ratios are 10\%-20\%.

The above result can be interpreted in two ways. Either the 
the bulk of \fe2\ arises in a dusty region,  or the
{\it A-}values for the transitions involved need a fine 
tuning. In the first case, it implies that the extinction
affecting the gas emitting the \fe2\ lines is large and 
surprisingly homogeneous. Assuming 
an intrinsic 1.257$\mu$m/1.644$\mu$m ratio of 1.34, the 
observed ratios agree with A$_{\mathrm{v}} \sim$3.7 
(adopting the Cardelli et al. \cite{ccm89} extinction law 
and foreground screen). This last value would be even 
higher in Mrk\,1239 and NGC~4051 but overall, rather similar 
in most AGN. Note that the extinction towards the NLR 
determined by other authors for some of the objects of our 
sample is significantly lower: A$_{\mathrm{v}} \sim$0.6 for 
NGC\,3227 (Crenshaw et al. \cite{cr01}), $\sim$0.43 for
Ark~564 (Crenshaw et al. \cite{cr02}) and $\sim$1.27 for Mrk~766 
(Gonz\'alez Delgado \& P\'erez \cite{gdp96}). 

In the second case, our data can be useful to set important
constraints to a more precise determination of the 
{\it A-}values for \fe2\ forbidden lines. Bautista \&
Pradhan (\cite{bp98}) had already noted about the
difficulty associated to the determination of accurate
forbidden transition probabilities, especially for
\ion{Fe}{ii}, owing to the large numbers of algebraic
terms involved. Unfortunately, there are very few
works published in the literature with a systematic 
determination of the ratio {\it I}(1.257$\mu$m/1.644$\mu$m)
on a similar sample. However, our results tend to agree to 
that of Sugai et al. (\cite{su99}), who
reports a ratio of 0.80 for ARP~299. The extinction A$_{\mathrm{v}}$,
determined by Heisler \& De Roberties
(\cite{hdr99}) from {\it I}(1.257$\mu$m/1.644$\mu$m) on a small
sample of \ion{H}{ii} and Seyfert 2 galaxies, also tends
to be significantly higher that that derived from optical
data.

In short, our observations imply that the extinction
measured from the \fe2\ lines is large compared to that
determined from optical indicators. This can be due
to a genuine effect and it is somehow expected because
the NIR probes a larger optical depth than the optical.
An alternative is to consider that the intrinsic
1.257$\mu$m/1.644$\mu$m ratio is $\sim$30\% smaller
than the actual theoretical value. This is certainly an issue
that deserves further attention and need to be 
explored using a larger sample of objects.

\section{Final Remarks}

In order to discuss the origin of the molecular hydrogen and
\fe2 emission lines, near infrared observations have been
carried out for a sample of 13 NLS1, 6 Seyfert~1, 2 Seyfert~2 and 1 
starburst galaxies. In the following, our
main conclusions are briefly reviewed.

The kinematical link between the molecular hydrogen and the NLR gas
was analysed on the basis of the line widths. For all galaxies in 
the sample, \h2\ 2.121$\mu$m was in the limit of the spectral resolution 
(360 \kms) and narrower than the NLR lines for the cases in which
the latter lines were resolved. \fe2\ was resolved for 50\% of the sample. 
In NGC\,3227, NGC\,4151 and PG\,1612+261 it was
broader than [\ion{S}{vi}] 1.963$\mu$m. It is then unlikely that \h2\ and \fe2\ 
come from the same volume of gas.

The detection of \h2\ (1,0)S(1) 2.121$\mu$m in 20 out of the 22 
objects of our sample reveals, for the first time, strong evidence 
that molecular gas within 500~pc from the 
centre is common in AGN and not restricted to SB/AGN composite galaxies.
\h2\ lines from vibrational transitions were detected but are not
as bright as in planetary nebulae, providing the first indication that
UV fluorescence is not the main mechanism that excites them.
This result is strengthened  by the analysis of the line ratios that
discriminates between shock-dominated and fluorescence production.
For 4 objects (NGC~3227, NGC~4051, NGC~4151 and
Mrk~766) the excitation mechanism is clearly thermal.
For the remaining AGN, a mixing with a non-thermal process cannot
be discarded, although the results point to a dominant thermal mechanism.
This result is confirmed by the similarity between the 
vibrational and rotational temperatures of the \h2. However, 
a unique source of the thermal component could not be established. The
relative contribution to this component coming from shocks originating
in SN explosions or in radio jets, and from X-rays heating may vary
from one object to the other.

The origin of the \fe2\ lines is also controversial. \fe2\
1.257$\mu$m was detected in all galaxies of our sample, suggesting
a link to the AGN itself. The \fe2\ 1.257$\mu$m/Pa$\beta$ ratio is
used to study the origin of the \fe2 emission. It reveals
that  stellar and non-stellar processes contributes to the \fe2\
spectrum with a predominance of the latter in most objects. The
\fe2\ 1.257$\mu$m/Pa$\beta$ ratio values are within the interval
usually observed in other AGN. Mrk~766 is particular in this respect,
however. Being a NLS1, the observed \fe2\ is compatible with a
starburst origin.

A  correlation between \h2\ 2.121$\mu$m/Br$\gamma$ and
\fe2\ 1.257$\mu$m/Pa$\beta$ was found but it
breaks down for the galaxies where either one of the ratios
is larger than 2, predominantly Seyfert~2s.
Starburst galaxies are preferentially located in the region
with \fe2/Pa$\beta$ and \h2/Br$\gamma <$ 0.4 while LINERS
are characterized by \fe2/Pa$\beta$ and \h2/Br$\gamma >$ 2.
We confirm the usefulness of this plot as a diagnostic
tool at separating emitting line objects by activity level
and set constraints to the locus of points shown by AGN.

Line intensities predicted by X-ray heating
models are able to explain the \h2\ and part of the \fe2\ emission
in Seyfert~1 but fail at reproducing these characteristics in
Seyfert~2.   The results are improved
if the emitting clouds are located closer
to the central source, $\sim$80~pc or less. This would explain
the correlation between \h2\ 2.121$\mu$m/Br$\gamma$ and
\fe2\ 1.257$\mu$m/Pa$\beta$  discussed above. However, the presence
of significant starburst activity  in some of the objects makes
it plausible that more than one exciting mechanism is at work.
In Seyfert~2,  X-rays appears to have a lesser
importance and  the \h2  emission is probably enhanced by  contributions 
from circumnuclear starburst and structures such as ionization cones 
and dust lanes. Nonetheless, the role of these structures is not
confirmed by the present data. 

Finally, our data allowed us to study, for the first time,
the \fe2\ 1.257$\mu$m/1.644$\mu$m ratio in Seyfert~1 galaxies.
We found that most objects have values 30\% lower than
the intrinsic value predicted by theory. It implies either
than the extinction towards the \fe2\ emitting clouds is
very similar in most objects and significantly larger
than that derived from optical observations, or there is
an overestimation of the {\it A}-values for the \ion{Fe}{ii}
transitions. We discarded a systematic error on the measurements
because the uncertainties associated to the observed ratio is
between 10\%-20\%.

\begin{acknowledgements}
This paper is partially supported by the Brazilian funding agencies
FAPESP (00/06695-0) and CNPq(304077/77-1) to ARA and SMV and the 
U.S. National Science Foundation (AKP). The authors thanks to
the comments and suggestions of an anonymous referee, which 
help to improve this paper. 
\end{acknowledgements}

\end{document}